\documentclass[12pt]{iopart}
\pdfoutput=1
\usepackage{iopams}  
\expandafter\let\csname equation*\endcsname\relax
  \expandafter\let\csname endequation*\endcsname\relax
\usepackage{amsmath}
\usepackage{graphicx,graphics,cite}
\usepackage{color}

\renewcommand{\v}{\mathbf{v}}

\newcommand{\x}{\mathbf{x}}
\newcommand{\mc}{\mathcal}

\begin{document}

\title[]{Persistent search in confined domains: a velocity-jump process model}

\author{Daniel B. Poll$^1$ and Zachary P. Kilpatrick$^{1}$}

\address{$^1$ Department of Mathematics, University of Houston, Houston TX}
\ead{zpkilpat@math.uh.edu}
\vspace{10pt}
\begin{indented}
\item[]date submitted
\end{indented}

\begin{abstract}
We analyze velocity-jump process models of persistent search for a single target on a bounded domain. The searcher proceeds along ballistic trajectories and is absorbed upon collision with the target boundary. When reaching the domain boundary, the searcher chooses a random direction for its new trajectory. For circular domains and targets, we can approximate the mean first passage time (MFPT) using a Markov chain approximation of the search process. Our analysis and numerical simulations reveal that the time to find the target decreases for targets closer to the domain boundary. When there is a small probability of direction-switching within the domain, we find the time to find the target decreases slightly with the turning probability. We also extend our exit time analysis to the case of partitioned domains, where there is a single target within one of multiple disjoint subdomains. Given an average time of transition between domains $\langle T \rangle$, we find that the optimal rate of transition that minimizes the time to find the target obeys $\beta_{{\rm min}} \propto 1/ \sqrt{\langle T \rangle}$.
\end{abstract}

%

\section{Introduction}


Organisms frequently search to find targets whose position is unknown to them. For example, animals search for food or mates in ways that balance both speed and low energy expenditure~\cite{charnov76,obrien90,viswanathan99}. In addition, the dynamics of biomolecules can be modeled as a search process. Recently, experiments and modeling studies have identified biochemical processes whose kinetics involve the search for a reaction partner, due to the small number of reactive molecules~\cite{berg81,vonhippel07,bressloff13}. Regardless of the context of searches, it is often desirable to minimize the amount of time needed to find the target, and this is the most common measure of search efficiency~\cite{benichou05}.

There are two particularly well studied models of random search processes: passive diffusion and intermittent search. Passive diffusion to a small target in a confined domain is a common model of molecular transport at the biomolecular scale~\cite{hanggi90,schuss07}. A distinct advantage of this framework is that the average time to find the target can be formulated as the solution to a mean first passage time (MFPT) problem~\cite{holcman04,condamin07,pillay10}. However, this model is not appropriate in all situations. In particular, foraging organisms and biomolecules that move ballistically often employ an intermittent search strategy, wherein diffusive search periods are punctuated by rapid displacement phases during which no search occurs~\cite{benichou11}. Such intermittent strategies can be optimized to obtain a minimal MFPT by balancing time spent in the moving and searching phases~\cite{loverdo08}.

In contrast to such previous work, one could also consider strategies wherein search is persistent and ballistic. The searcher then proceeds according to a velocity-jump process, moving ballistically and then switching direction at random times~\cite{othmer88,codling08}. The diffusion limits of velocity-jump processes are given by linear transport systems, specific cases of the Boltzmann equation~\cite{case67,othmer00}. Recently, this model has been used to analyze the statistics of foraging insect movement~\cite{grunbaum98,khuong13}. Presuming an animal can detect targets while moving ballistically~\cite{reynolds09}, search and travel can be modeled as a single process. One well studied experimental paradigm wherein an animal searches persistently is the Morris water navigation task, in which a rodent must locate a platform in a circular pool~\cite{vorhees06}. Visual search in psychophysics tasks is another example of persistent search, where the focal point of gaze moves ballistically in search of a visual target~\cite{schall99,rayner09}. Thus, concrete quantitative models are needed to understand the dynamics of persistent search and identify optimal strategies.

We analyze an idealized model of persistent search, which considers movements of the searcher to be ballistic trajectories with constant speed. For simplicity, we consider two-dimensional circular domains with reflecting boundaries along with circular targets with absorbing boundaries. Initially, we develop an asymptotic theory for approximating the time to find the target when the searcher only turns when encountering the domain boundary. This allows us to understand how the placement of the target impacts the average time to locate it. We extend our analysis to the case where the searcher turns on the domain interior with finite probability, showing this decreases the MFPT for low-probability of turning. Lastly, we introduce a model of persistent search on multiple disjoint domains. When the transit time between subdomains is nonzero, there is an optimal rate of transition between domains that balances domain coverage with the time penalty for traveling between domains. In all cases, we identify how search and domain parameters impact the MFPT.

\section{Velocity-jump process model of persistent search}
\label{model}
\begin{figure}[t]
\begin{center}
\includegraphics[width=6cm]{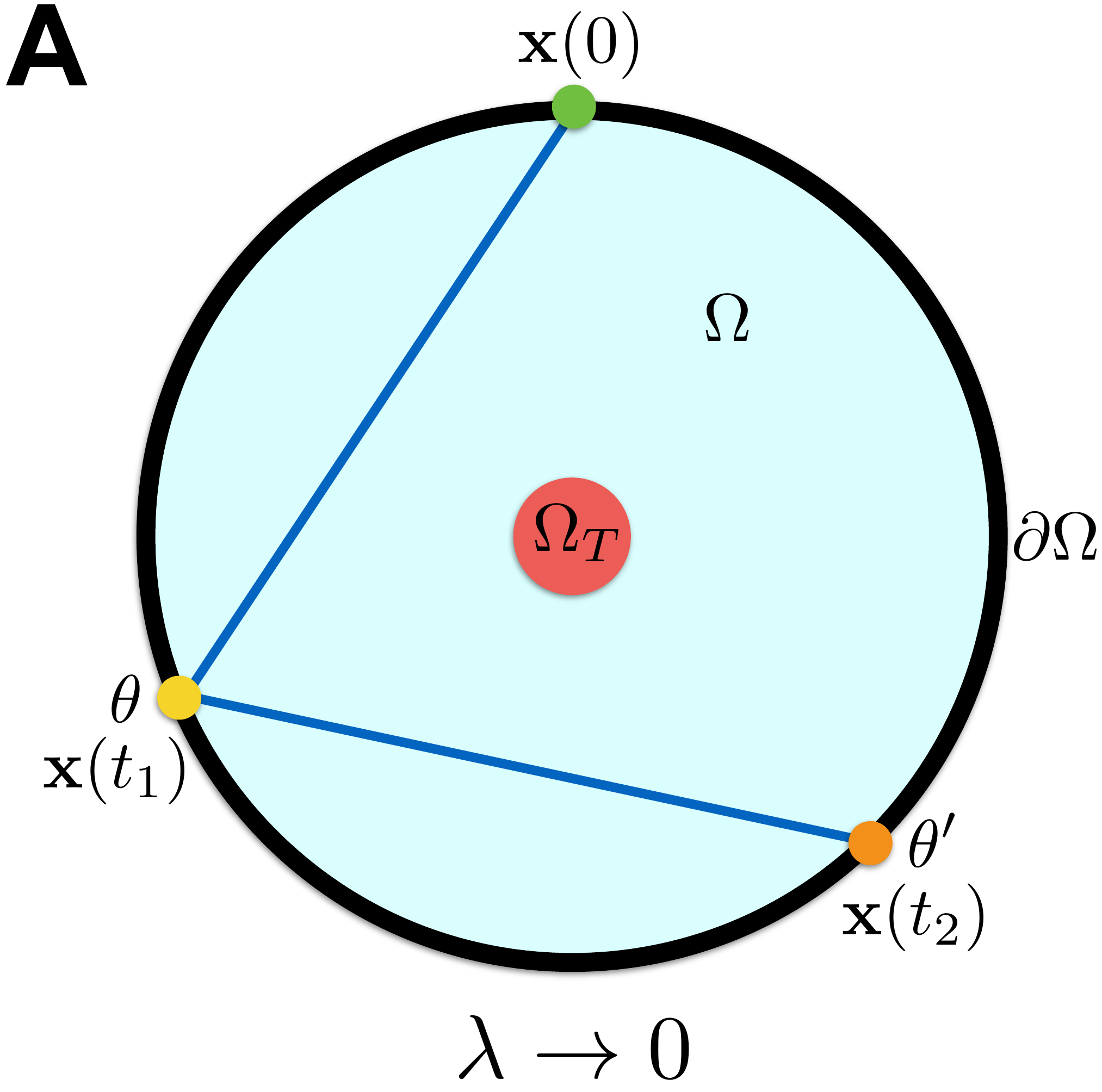} \hspace{5mm} \includegraphics[width=6cm]{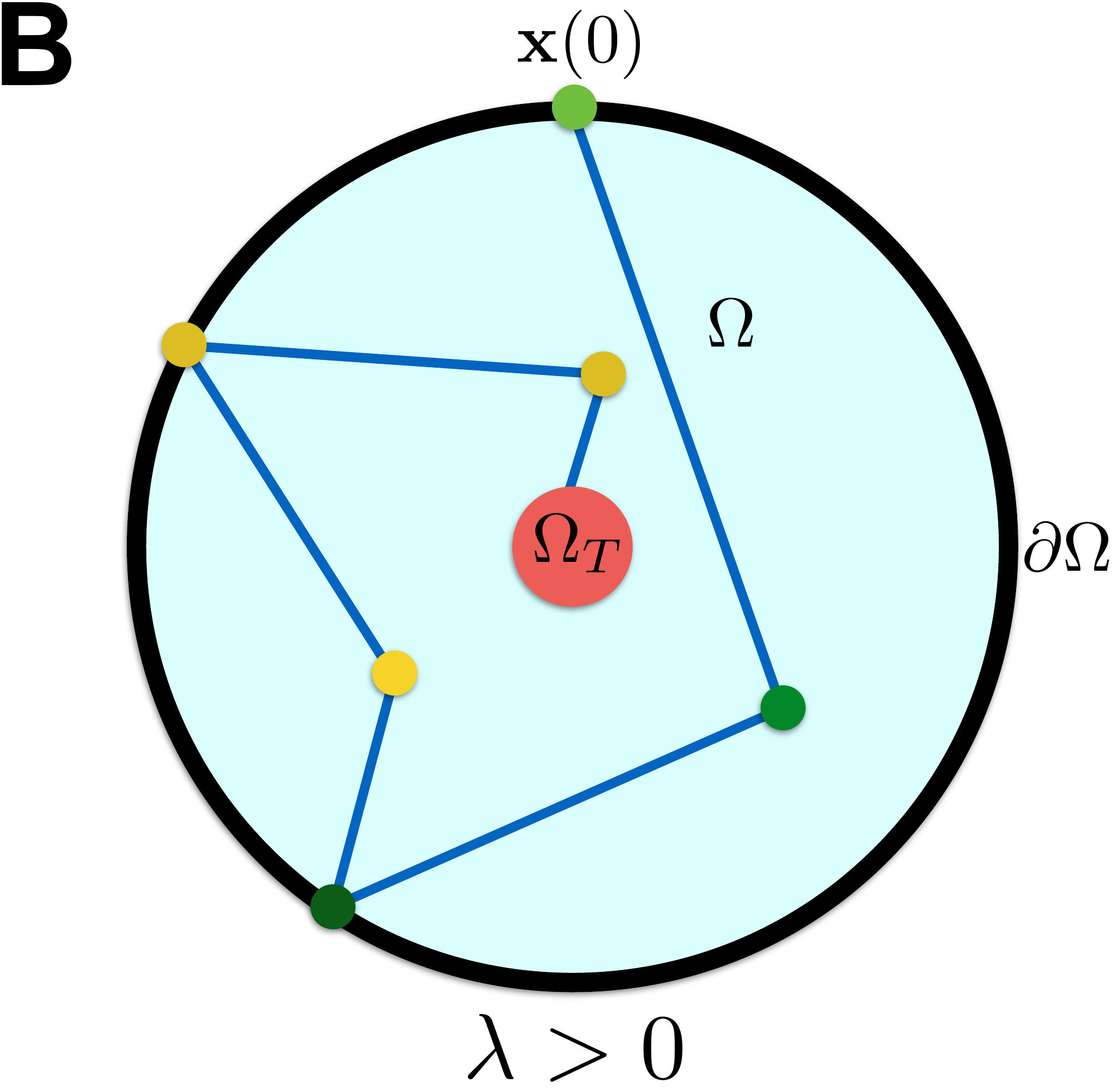} \end{center}
\caption{Velocity-jump process model of a persistent searcher in a confined domain. ({\bf A}) In the absence of interior turning ($\lambda \to 0$), the searcher moves ballistically between locations on the boundary $\partial \Omega$. A subsequent boundary location $\theta'$ is selected based on the current one $\theta$ by drawing from the probability density function $f(\theta - \theta')$. ({\bf B}) In the case of interior turning ($\lambda >0$), the searcher's trajectories are no longer wholly determined by the deflection angles at the boundary. The search concludes when the target domain $\Omega_T$ is encountered.}
\label{modelscheme}
\end{figure}
Consider the following model for the stochastically evolving position $\x (t)$ of a persistent searcher. We construct a model of a particle searching for a hidden target in a bounded, circular domain $\Omega$ of radius $R$, i.e. $\Omega := \big\{ (x,y) \in \mathbb{R}^2 : \sqrt{ x^2 + y^2 } \leq R \big\}$. The hidden target is also defined by a circular region with radius $r$:
\begin{align*}
\Omega_T := \big\{ (x,y) \in \Omega : \sqrt{ (x-x_0)^2 + (y-y_0)^2 } \leq r \big\},
\end{align*}
where $(x_0,y_0)$ denotes the centroid of the target domain. Note, we will restrict $x_0^2 + y_0^2 \leq R - r$, so the entire target is contained in the domain $\Omega$. Furthermore, due to the rotation symmetry of the circular domain, we exclusively consider targets with coordinates along the right horizontal radius, $(x_0, y_0) = ( \epsilon R, 0)$. All other cases can be reduced to this form by an axial rotation. 

The searcher's position evolves according to a velocity-jump process~\cite{othmer88}. On the interior of the domain $\x (t) \in \Omega \backslash \partial \Omega$, the searcher moves ballistically with velocity $\v (\phi) = v ( \cos \phi, \sin \phi )$ with constant speed $v$ and angle $\phi \in [0, 2\pi)$. Transitions in the velocity angle $\phi$ are governed by a continuous-time Markov process with turning rate $\lambda$. In the limit $R \to \infty$, the distribution of ballistic path-lengths is $p( v \cdot \Delta t) = \lambda \e^{- \lambda v \cdot \Delta t}$. Note that as $\lambda \to \infty$, $p(v \cdot \Delta t) \to  \delta(v \cdot \Delta t)$ and the searcher will exhibit Brownian motion~\cite{othmer88}. In the other extreme, $\lambda \rightarrow 0$, the searcher's probability of turning in finite time decreases to zero, maintaining its initial velocity $\v (\phi)$ until encountering the boundary $\partial \Omega$.

On the domain boundary $\x (t) \in \partial \Omega$, the searcher uses a separate rule for turning. For simplicity, we denote the searcher's position on the boundary according to its angle $\theta$ on the circular domain boundary $\x (t) = R ( \cos \theta, \sin \theta)$. It then selects a new heading direction by drawing a random variable $\theta'$ for the new boundary location it will move towards (Fig. \ref{modelscheme}{\bf A}). This random variable is chosen from the probability density function $f(\theta - \theta')$, an even symmetric function of the argument $\theta - \theta'$. In particular, this is a probability density function over $\theta'$, so that $\int_0^{2 \pi} f( \theta - \theta') d \theta' = 1$ for all $\theta \in [0, 2 \pi)$. A search ends once the target has been hit, which occurs when the searcher encounters the absorbing target boundary $\partial \Omega_T $ (Fig. \ref{modelscheme}{\bf B}).

An alternate description of searcher motion is studied in Section \ref{spiralsect}, where we consider spiral paths of motion into and out of the domain center. The MFPT for both spiral path and random ballistic path strategies are compared therein. Furthermore, we will consider extensions of this single domain model in Section \ref{multidomain}, when we incorporate movements of the searcher between multiple subdomains. In this case, movement on the subdomain interiors will proceed as before, but encounters with the boundary can lead to switches between subdomains.



\section{Purely ballistic search in single domains}

We begin by analyzing the model in the case of no turning on the domain interior ($\lambda \to 0$). The searcher proceeds from one point on the domain boundary to another along straight trajectories, unless it encounters the target domain $\Omega_T$. For a trajectory from the domain location $\theta$ to $\theta'$, there is probability $a(\theta,\theta')$ of passing through the target domain $\Omega_T$. Since a collision with the target always results in absorption, this means $a(\theta, \theta')$ is an indicator function, equaling 1 if the trajectory from $\theta$ to $\theta'$ passes through the target and 0 otherwise. Marginalizing over all possible paths, we can compute average probability of passing through $\Omega_T$, which we define as $\bar{a}$. To compute $\bar{a}$, we integrate:
\begin{align}
\label{ava}
\bar{a} = \frac{1}{4 \pi^2 }\int\limits_0^{2\pi} \int\limits_0^{2 \pi} a(\theta,\theta') f(\theta - \theta') d \theta d\theta'.
\end{align}
The boundary conditions of the velocity-jump process on $ \partial \Omega$ imply that $a(\theta,\theta')$ is weighted by the probability of sampling $\theta'$ given $\theta$, defined to be $f(\theta,\theta')$. The domain is radially symmetric, and we assume a uniformly random initial conditions $\x (0)$ along the boundary $\partial \Omega$. Subsequent angles $\theta$ could be non-uniformly distributed on the boundary, but this does not substantially impact our approximations.

Typically, the limiting quantity in search problems is the average search time~\cite{benichou11}. We start by computing the average time of a single path from one boundary location to another. Without loss of generality, we rescale our domain and target sizes ($R$ and $r$) so the constant search velocity is $v=1$ in new coordinates. Thus, we need only find the average distance of a path by calculating the chord length of a circle of radius $R$ from angle $\theta$ to $\theta'$, given by ${\rm ch}= 2R \sin({\rm mod}(\theta - \theta', 2 \pi)/2)$. 
Assuming uniform distributions of angle locations $\theta$, the average chord length of each path will be
\begin{align}
\label{avch}
\overline{{\rm ch}} = \frac{R}{\pi} \int_0^{2\pi}  \sin\Big( \frac{z}{2} \Big) f(z) d z,
\end{align}
where we have applied the change of variables $\theta - \theta' \mapsto z$.

\subsection{Small and centered targets}

We can utilize Eqs.~(\ref{ava}) and (\ref{avch}) to help us determine the average time it will take until the searcher reaches the target domain $\Omega_T$.  Our approximation assumes the target absorption probability $\overline{a}$ is fixed across all paths. In the limit of small target sizes $r \ll R$, the distribution of boundary locations over $\theta$ will remain near uniform after the searcher makes a single path between trajectories.
With this in mind, we can compute the average time to hit the target by treating absorption into the target as a Markov process. Let $h_j = \bar{a}(1-\bar{a})^j$ denote the probability of hitting the escape domain $\Omega_T$ on the $j^{th}$ path across the domain, and let $d_j = j \cdot \overline{{\rm ch}} + R$ denote the average distance of all paths to the target that end on the $j^{th}$ pass.
We can compute the MFPT for the absorption of the searcher into the target using the weighted sum:
\begin{align}
\label{binapp}
{\mc T} = \sum_{j=0}^\infty h_j d_j = \bar{a} \sum_{j=0}^{\infty} (1 - \overline{a})^j (j \cdot \overline{{\rm ch}} + R) =   \overline{{\rm ch}} \Big( \frac{1}{\overline{a}} - 1 \Big)  + R.
\end{align}
Note, the time to find the target increases as the likelihood of finding it in a single path $\overline{a}$ decreases. Furthermore, increasing the size of the domain $R$ increases the time to find the target via $\overline{{\rm ch}}$ and $\bar{a}$. We can gain further insight by studying the effects of target size on the probability $\overline{a}$.

\begin{figure}[t]
\begin{center} \includegraphics[width=5cm]{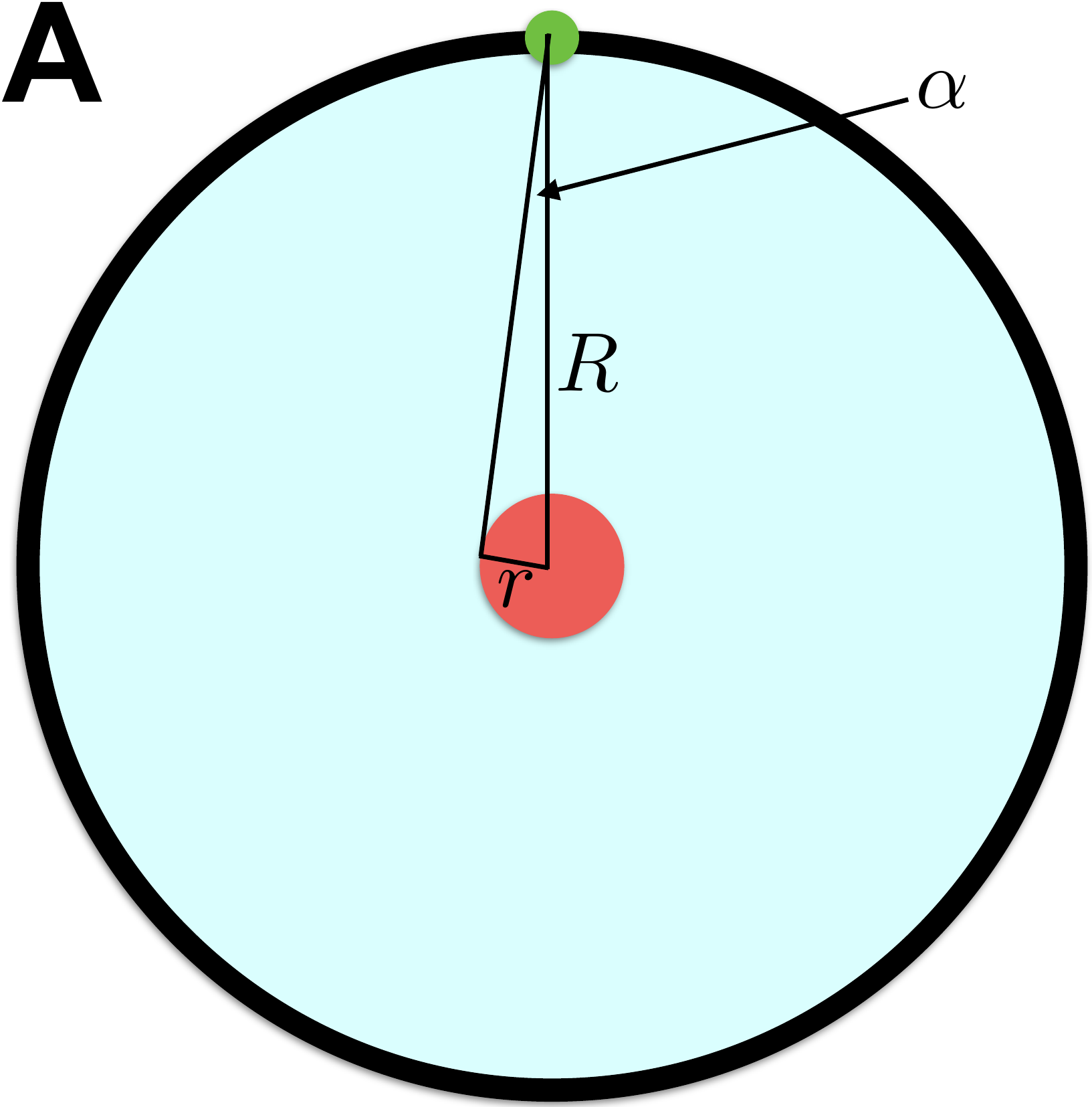} \hspace{10mm}
\includegraphics[width=8cm]{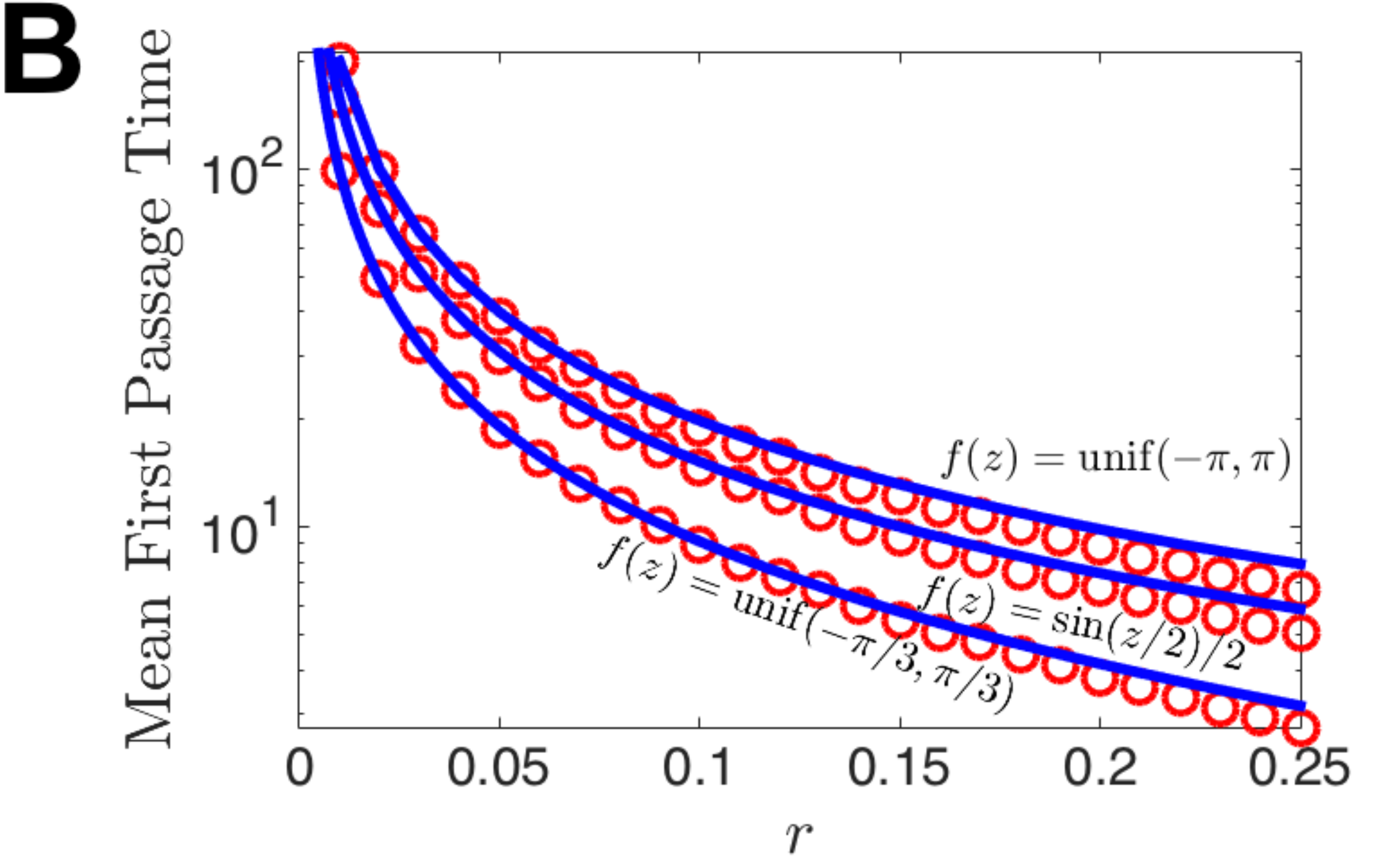}
\end{center}
\caption{Ballistic search for a small target in the domain center. (A) The set of trajectories that hit the target from the boundary are contained in the arclength $2 \alpha$, where $\alpha = \sin^{-1} (r/R)$. (B) MFPTs for different forms of the probability density $f(\theta - \theta')$. Results for uniform densities $f(z) = {\rm unif}[- \gamma,\gamma]$ are shown as well as for the nonconstant function $f(z) = \sin (z/2) / 2$. The domain radius is $R=1$. Each circle is the MFPT from $10^6$ numerical simulations of the process in Section \ref{model} with $\lambda = 0$.}
\label{centertarget}
\end{figure}

By fixing the target in the center of the domain, the probability of hitting the target is the same for any boundary location, so $a(\theta, \theta')$ only depends on the difference $\theta - \theta'$ between the angles of a path's endpoints. In particular, the searcher is absorbed in a region of arc length $2 \alpha$, so $a(\theta - \theta') = 1$ if $|\theta - \theta'| \leq \alpha$ and $a(\theta - \theta') = 0$ otherwise (Fig. \ref{centertarget}A). We also assume a new search direction is chosen uniformly from a symmetric region oriented toward the center, $f(\theta - \theta') = \frac{1}{2 \gamma}$ if $| \theta + \pi  - \theta' |  \leq \gamma $ and $f(\theta - \theta') =0$ otherwise. This simplifies the average hitting probability $\overline{a}$ as follows
\begin{align*}
\overline{a} = \frac{1}{2 \pi}\int\limits_0^{2\pi} \int\limits_0^{2 \pi} a(\theta - \theta') f(\theta - \theta') d \theta d\theta'  = \frac{1}{4 \pi \gamma} \int_0^{2 \pi} \int_{\pi - \alpha}^{\pi + \alpha}  d z d \theta =   \frac{\alpha}{\gamma}
\end{align*}
for $\gamma > \alpha$. Since the target is small ($r \ll R$), we expect $\alpha \ll \gamma$. The arclength $2 \alpha$ can be found using trigonometry to be $2 \sin^{-1} (r/ R)$, so that $\overline{a} = 2 \sin^{-1} (r/R) / \gamma$.

Furthermore, the average chord length $\overline{{\rm ch}}$ is computed by marginalizing against the probability density function of paths from $\theta$ to $\theta'$, yielding
\begin{align}
\overline{{\rm ch}} = \frac{1}{2 \pi} \int\limits_0^{2\pi} \int\limits_{0}^{2 \pi} 2R \sin\left( \frac{\theta - \theta'}{2} \right) f(\theta - \theta') d \theta' d \theta = 
\frac{R}{\gamma} \int\limits_{\pi - \gamma}^{\pi + \gamma} \sin \frac{\theta}{2} d \theta = \frac{4R \sin(\gamma/2)}{\gamma}. \label{chordgamma}
\end{align}
Fixing the target in the center of the domain $\Omega$ and choosing a uniform distribution for the density $f(\theta - \theta')$  produces an approximate MFPT as
\begin{align}
{\mc T}_c = \frac{4R \sin(\gamma/2)}{\gamma} \bigg( \frac{\gamma}{2 \sin^{-1}(r/R)} - 1 \bigg) + R, \label{mfptcenter}
\end{align}
which matches well with numerical simulations of the velocity-jump process (Fig. \ref{centertarget}B). For a smaller search arclength $\gamma$, the MFPT decreases, since there is a higher probability of heading toward the target $\Omega_T$ from the boundary $\partial \Omega$.

We also demonstrate that Eq.~(\ref{mfptcenter}) is monotone in each parameter by taking partial derivatives. First, we show the MFPT decreases with the target radius $r$. Differentiating Eq.~(\ref{mfptcenter}) with respect to $r$ yields
\begin{align*}
\frac{\partial {\mc T}_c}{\partial r} = - \frac{2 R \sin (\gamma/2)}{ \sqrt{R^2 - r^2} \cdot \left[ \sin^{-1} (r/R) \right]^2} < 0,
\end{align*}
so larger targets are found faster. Second, note that
\begin{align*}
\frac{\partial {\mc T}_c}{\partial \gamma} = \frac{R \cos (\gamma/ 2)}{\sin^{-1} (r/R)} - \frac{2R \left[ \gamma \cos (\gamma/2) - 2 \sin (\gamma/2) \right]}{\gamma^2}, 
\end{align*}
so when $\gamma = \pi$, we have
\begin{align*}
\left. \frac{\partial {\mc T}_c}{\partial \gamma}  \right|_{\gamma = \pi} = \frac{4 R}{\pi^2},
\end{align*}
and the MFPT decreases as $\gamma$ is decreased from $\pi$ (Fig. \ref{centertarget}B). Finally, we can show that the MFPT increases as the domain radius $R$ is increased by differentiating
\begin{align*}
\frac{\partial {\mc T}_c}{\partial R} = - \frac{4 \sin( \gamma/2)}{\gamma} +  2\sin (\gamma/2) \left[  \frac{r + \sqrt{R^2 - r^2} \sin^{-1}(r/R)}{\sqrt{R^2 - r^2} \cdot \left[ \sin^{-1}(r/R) \right]^2} \right]  +1, 
\end{align*}
so if we plug in $\gamma = \pi$, we find
\begin{align*}
\left. \frac{\partial {\mc T}_c}{\partial R} \right|_{\gamma = \pi} = 1 - \frac{4}{\pi} +  2 \left[   \frac{r + \sqrt{R^2 - r^2} \sin^{-1}(r/R)}{\sqrt{R^2 - r^2} \cdot \left[ \sin^{-1}(r/R) \right]^2} \right] >  1 - \frac{4}{\pi} + \frac{2}{\sin^{-1} (r/R)} >  1,
\end{align*}
since $R>r$, so $1/\sin^{-1} (r/R) > 2/ \pi$.

Now, considering the searcher may have an increased likelihood of searching toward the domain interior, we examine $f(\theta - \theta') = \sin((\theta - \theta')/2)/4$, which peaks at $\theta - \theta' = \pi$, the center of the domain. The average probability of hitting the target is then
\begin{align*}
\overline{a} = \frac{1}{8 \pi} \int_0^{2 \pi} \int_0^{2\pi} a(z) \sin \frac{z}{2} d z d \theta = \frac{1}{4}  \int_{\pi - \alpha}^{\pi + \alpha} \sin \frac{z}{2} d z = \sin \left[ \sin^{-1} (r/R) \right] = \frac{r}{R}.
\end{align*}
Furthermore, the average chord length is
\begin{align*}
\overline{{\rm ch}} = \frac{1}{8 \pi} \int_0^{2 \pi} \int_0^{2 \pi} 2 R \sin^2 \left( \frac{ \theta - \theta'}{2} \right) d \theta' d \theta = \frac{R}{2} \int_0^{2 \pi} \sin^2 \frac{z}{2} d z = \frac{R \pi}{2}.
\end{align*}
Applying Eq.~(\ref{binapp}) for the MFPT, we have
\begin{align}
{\mc T}_c = \frac{R \pi}{2} \left( \frac{R}{r} - 1 \right) + R.  \label{centersine}
\end{align}
As before, Eq.~(\ref{centersine}) is monotone decreasing in $r$ since $ \displaystyle \frac{\partial {\mc T}_c}{\partial r} = - \frac{\pi R^2}{2 r^2}$, whereas it is increasing in $R$ since $ \displaystyle \frac{\partial {\mc T}_c}{\partial R} = \frac{\pi R}{r} + 1 - \frac{\pi}{2}$ and $R>r$ (Fig. \ref{centertarget}B). Thus, our approximation using the expansion in Eq.~(\ref{binapp}) shows us explicitly that the MFPT decreases with target size $r$ and increases with domain size $R$. In addition, we note the MFPT approximations can be further truncated to the asymptotic form ${\mc T}_c = {\mc C}/r$ (with constant ${\mc C}$) in the $r \ll R$ limit, which still provides the appropriate form observed in Fig. \ref{centertarget}B.


\begin{figure}[t]
\begin{center} \includegraphics[width=5cm]{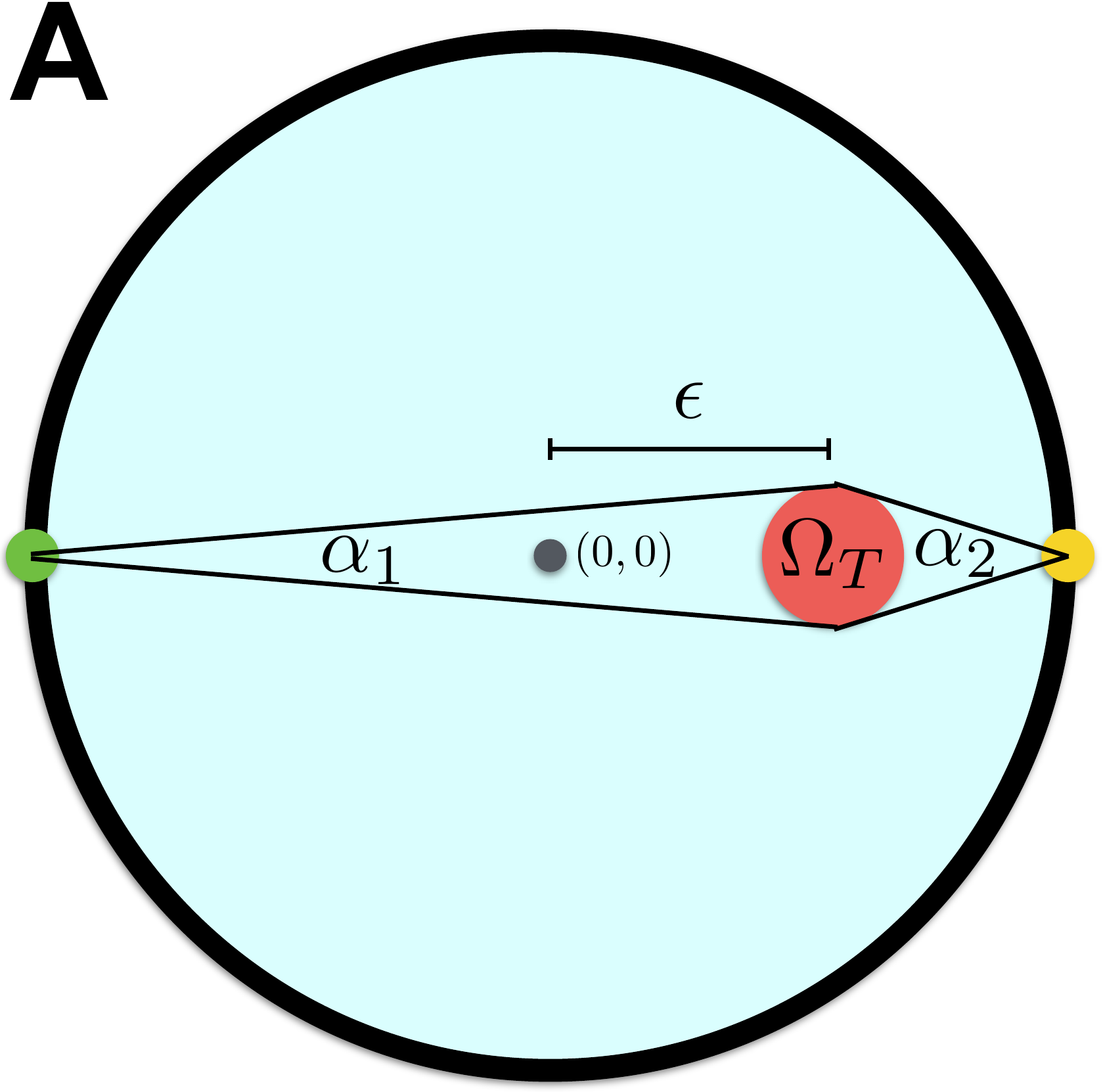} \hspace{8mm}
\includegraphics[width=8cm]{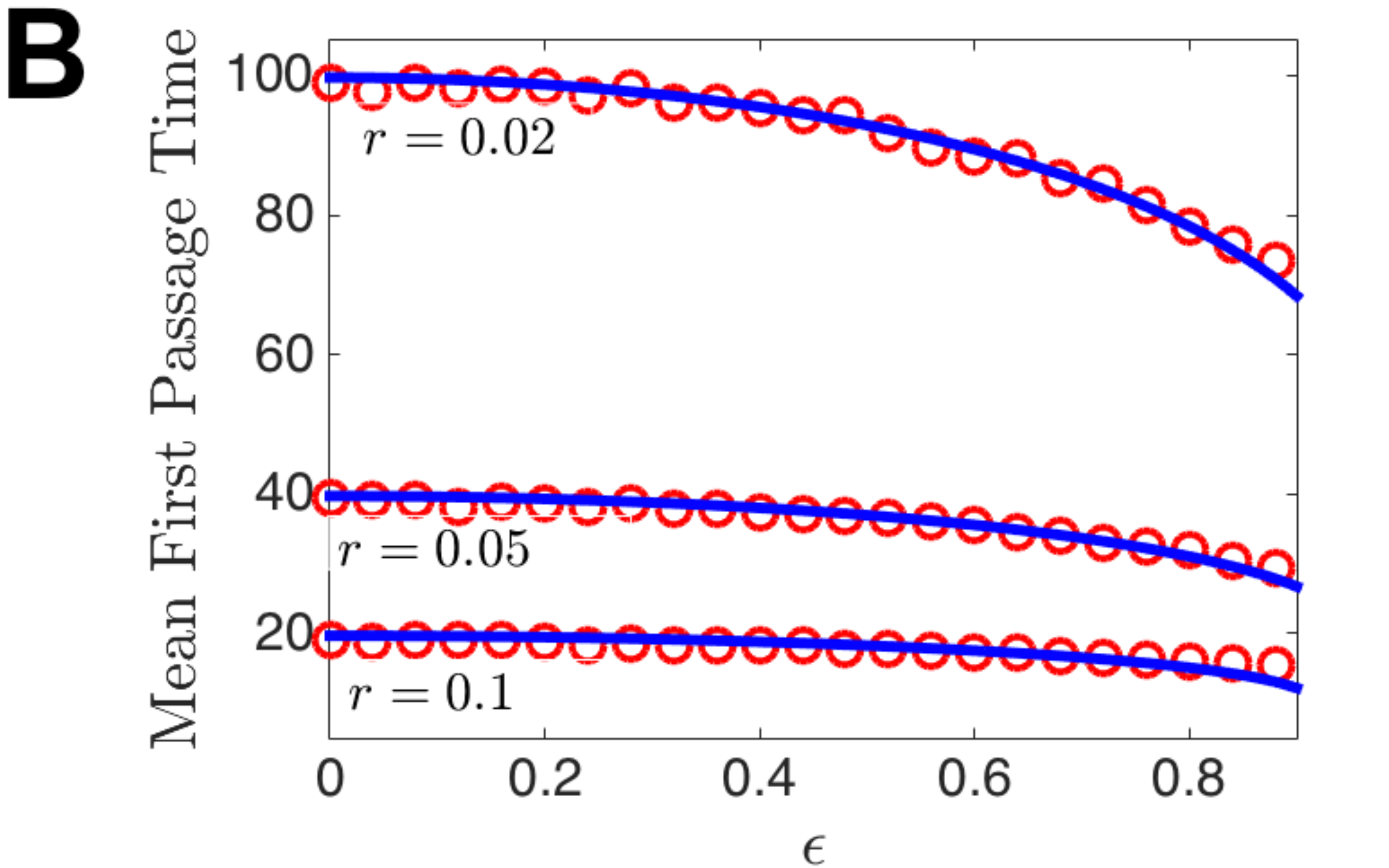} \end{center} 
\caption{MFPT decreases with the eccentricity $\epsilon$ of the target. (A) Off-center targets are shifted to location $(x_0,y_0)=(\epsilon R, 0)$. The target angle $\alpha_1$ is smaller on the far side of the boundary than on the close side $\alpha_2$.  (B) MFPT as a function of the eccentricity $\epsilon$ comparing theory (solid line) using hitting probability $\bar{a}(\epsilon)$ given by Eq.~(\ref{targaoff}) to numerical simulations (circles). The angle of search from the boundary is drawn from the density function $f(z) = {\rm unif}[- \pi, \pi]$. }
\label{offcenter}
\end{figure}

\subsection{Off-center Targets}

We now study the time to hit off-center targets ($(x_0,y_0) = (\epsilon R, 0)$). Interestingly, we find the MFPT decreases as the target nears the domain boundary $\partial \Omega$ (as $\epsilon$ increases). In this case, the  discrepancy between the largest and smallest angle from the boundary containing the target grows with $\epsilon$ (Fig. \ref{offcenter}A). Nevertheless, we can marginalize and frame the target finding problem in terms of a Markov chain with constant probability of absorption, across realizations. 
However, due to the asymmetry introduced by placing the target away from the center, the indicator function $a(\theta, \theta')$ now depends on both variables independently. The angle containing the target can be derived by applying the law of sines to the triangle with: (a) one leg being the radius $R$ from the center to the boundary angle $\theta$; (b) another leg being the segment from the domain center to the target center, having length $\epsilon R$; and (c) the segment connecting (a) and (b). The angle $\psi$ of the triangle emanating from the boundary is then given by
\begin{align}
\frac{L(\theta)}{\sin \theta} = \frac{\epsilon R}{\sin \psi} = \frac{R}{\sin (\psi + \theta)},  \label{offlawsines}
\end{align}
where $L(\theta)$ is the length of the leg described by (c). Solving the system Eq.~(\ref{offlawsines}), we find that $\sin^2 \psi = \epsilon^2 \sin^2 \theta/ (1 - 2 \epsilon \cos \theta + \epsilon^2)$ and $L( \theta) = R \sqrt{1 - 2 \epsilon \cos \theta + \epsilon^2}$. As before, we use the leg length $L(\theta)$ to compute the target angle $\alpha (\theta) = \sin^{-1} ( r/ L(\theta))$ from the angle $\theta$ on the boundary. For boundary search direction distribution $f( \theta - \theta') = {\rm unif}[- \pi, \pi]]$, the average probability of finding the target with each path is approximated
\begin{align}
\overline{a} (\epsilon) &= \frac{1}{4 \pi^2} \int_0^{2 \pi} \int_0^{2 \pi} 2 \sin^{-1} \left[ \frac{r}{R \sqrt{1 - 2 \epsilon \cos \theta + \epsilon^2}} \right] d \theta' d \theta \nonumber \\
&= \frac{1}{ \pi} \int_0^{2 \pi} \sin^{-1} \left[ \frac{r}{R \sqrt{1 - 2 \epsilon \cos \theta + \epsilon^2}} \right] d \theta.  \label{targaoff}
\end{align}
Eq.~(\ref{targaoff}) converges to $\overline{a} (0) = 2 \sin^{-1} (r/R)/ \pi$ in the limit $\epsilon \to 0$. In the limit of small targets $r \ll R$, the average chord length is approximately same as in the center target case, Eq.~(\ref{chordgamma}). Thus, we can use $\overline{{\rm ch}} = 4 R/ \pi$. Plugging these Eqs.~(\ref{chordgamma}) and (\ref{targaoff}) into the asymptotic approximation Eq.~(\ref{binapp}) yields our asymptotic approximation (Fig. \ref{offcenter}B).

For small $\epsilon$, we can show the hitting probability $\overline{a}( \epsilon)$ increases with $\epsilon$, implying the MFPT decrease with $0< \epsilon \ll 1$ (Fig. \ref{offcenter}B). Note, $\overline{a}(\epsilon)$ is an even function, due to the circular symmetry of $\Omega$. Therefore, the ${\mc O}(\epsilon^2)$ term is the first term in a regular perturbation beyond $\overline{a}(0)$. Taking the second derivative at $\epsilon = 0$ yields
\begin{align}
\left. \frac{\partial^2 \bar{a}(\epsilon)}{\partial \epsilon^2} \right|_{\epsilon = 0} & = \frac{1}{\pi}  \int_0^{2 \pi} \frac{ \displaystyle \frac{r}{R} \left[ 1 + \left( 3 - 2 \frac{r^2}{R^2} \right) \cos (2 \theta) \right]}{\displaystyle 2 \left( 1 - \frac{r^2}{R^2} \right)^{3/2}} d \theta = \frac{r}{\displaystyle 2 \pi R \left( 1 - \frac{r^2}{R^2} \right)^{3/2}} . \label{abareps}
\end{align}
Since $r < R$, $\bar{a}''(0) > 0$, so the probability $\overline{a}(\epsilon)$ will increase with $\epsilon$ when $0 < \epsilon \ll 1$. Furthermore, we can write the asymptotic approximation
\begin{align*}
\overline{a}(\epsilon) = \frac{2}{\pi} \sin^{-1} \frac{r}{R} +  \frac{r \epsilon^2}{2 \pi R \left( 1 - \frac{r^2}{R^2} \right)^{3/2}} + {\mc O}(\epsilon^4).
\end{align*}
In sum, we have shown that a binomial expansion in hitting probabilities per path provide a reasonably accurate approximation of the MFPT for a purely ballistic search process. In the next section, we extend these results to account for the case of a searcher that makes turns on the interior of the domain $\Omega$.

\section{Interior turning in single domains}

\subsection{Turning via velocity-jumps}

We now explore how turning on the interior of the domain $\Omega$ affects the average time to find the target. Thus far, we have only considered searchers that turn on the boundary $\partial \Omega$. Interior turning is now incorporated according to a velocity-jump process. For infinitesimal timesteps $dt$, the probability of a velocity-direction change between $t$ and $t + dt$ is $\lambda \cdot dt$. Velocity changes are sampled from a uniform distribution so that the probability of selecting a new velocity with angle $\phi \in [0, 2 \pi)$ is ${\rm Pr}(\phi) = \frac{1}{2 \pi}$. For an unbounded domain ($R \to \infty$), this would lead to trajectories made of ballistic step-lengths $x$ over the distribution $p(x) = \lambda \e^{- \lambda x} $ for normalized velocity $|| \mathbf{v} || = 1$.

For low turning probability $\lambda \ll 1$, we asymptotically approximate the hitting probability for a single path between boundary points $\bar{a}( \lambda)$. Such paths are no longer necessarily comprised of a single straight segment; paths can be made up of two or more straight segments. However, we only focus on the change in hitting probability arising due to incorporating paths made of two straight segments. To begin, note the probability of not turning (number of turns $n=0$) along a segment of length $l$ is given
\begin{align}
{\rm Pr}( n = 0 | l) = 1 - \lambda \int_0^{l} \e^{- \lambda x} d x = \e^{- \lambda \cdot l}, \label{noturn}
\end{align}
so a searcher heading towards the target will not turn with approximate probability $\e^{- \lambda (R-r)}$. Thus, the likelihood that the searcher is absorbed into the target by following a single segment from the boundary is
\begin{align}
{\rm Pr}( {\rm hit} | n=0) {\rm Pr}( n = 0 | R-r) =  \e^{- \lambda (R-r)} \frac{2 \sin^{-1} (r/R)}{\pi}, \label{hitnoturn}
\end{align}
where we assume $f(z) = {\rm unif}[-\pi,\pi]$.

Next, we approximate the likelihood that the searcher is absorbed into the target by following two segments connected by a single turn. The likelihood of making at least one turn before hitting the boundary is given by subtracting the survival probability over the average chord length $\bar{\rm ch}$ from 1:
\begin{align}
{\rm Pr}( n > 0 | \bar{\rm ch}) = \lambda \int_0^{\bar{\rm ch}} \e^{- \lambda x} d x = 1 - \e^{- \lambda \cdot \bar{\rm ch}}. \label{oneturn}
\end{align}
Note also that the likelihood of more than one turn is ${\mc O}(\lambda^2)$ when $\lambda \ll 1$, since
\begin{align*}
{\rm Pr}(n > 1| x_1, x_2) &= \prod_{j=1}^2 \left( \lambda \int_0^{x_j} \e^{- \lambda y} d y \right) = \prod_{j=1}^2 \left( 1 - \e^{- \lambda x_j} \right) \approx \lambda^2 x_1 x_2 + {\mc O}(\lambda^3),
\end{align*}
where $x_1$ and $x_2$ are the maximal lengths of each segment. Therefore, we approximate the probability of there being a single turn by Eq.~(\ref{oneturn}).

\begin{figure}[t]
\begin{center} \includegraphics[width=5cm]{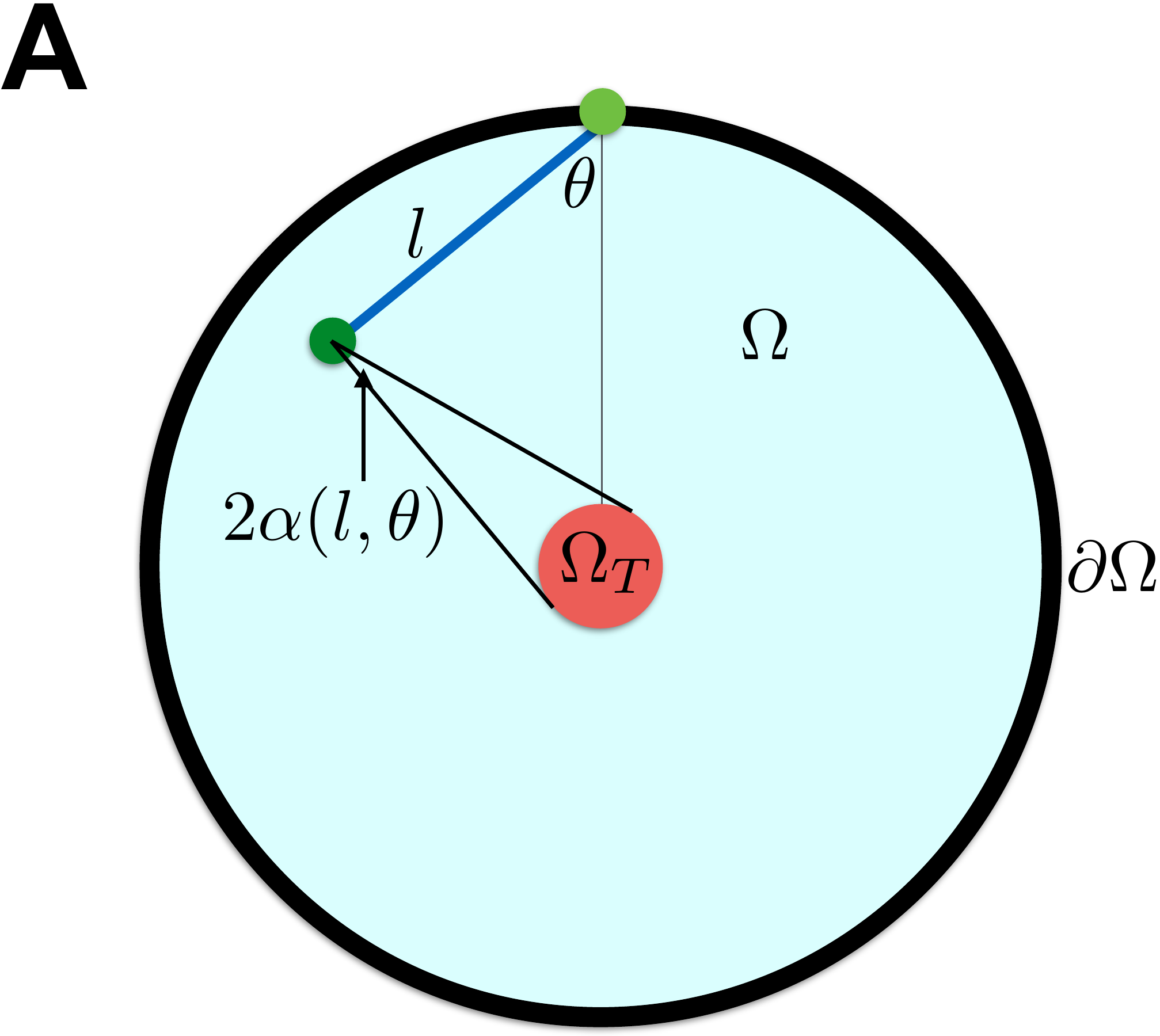} \includegraphics[width=6cm]{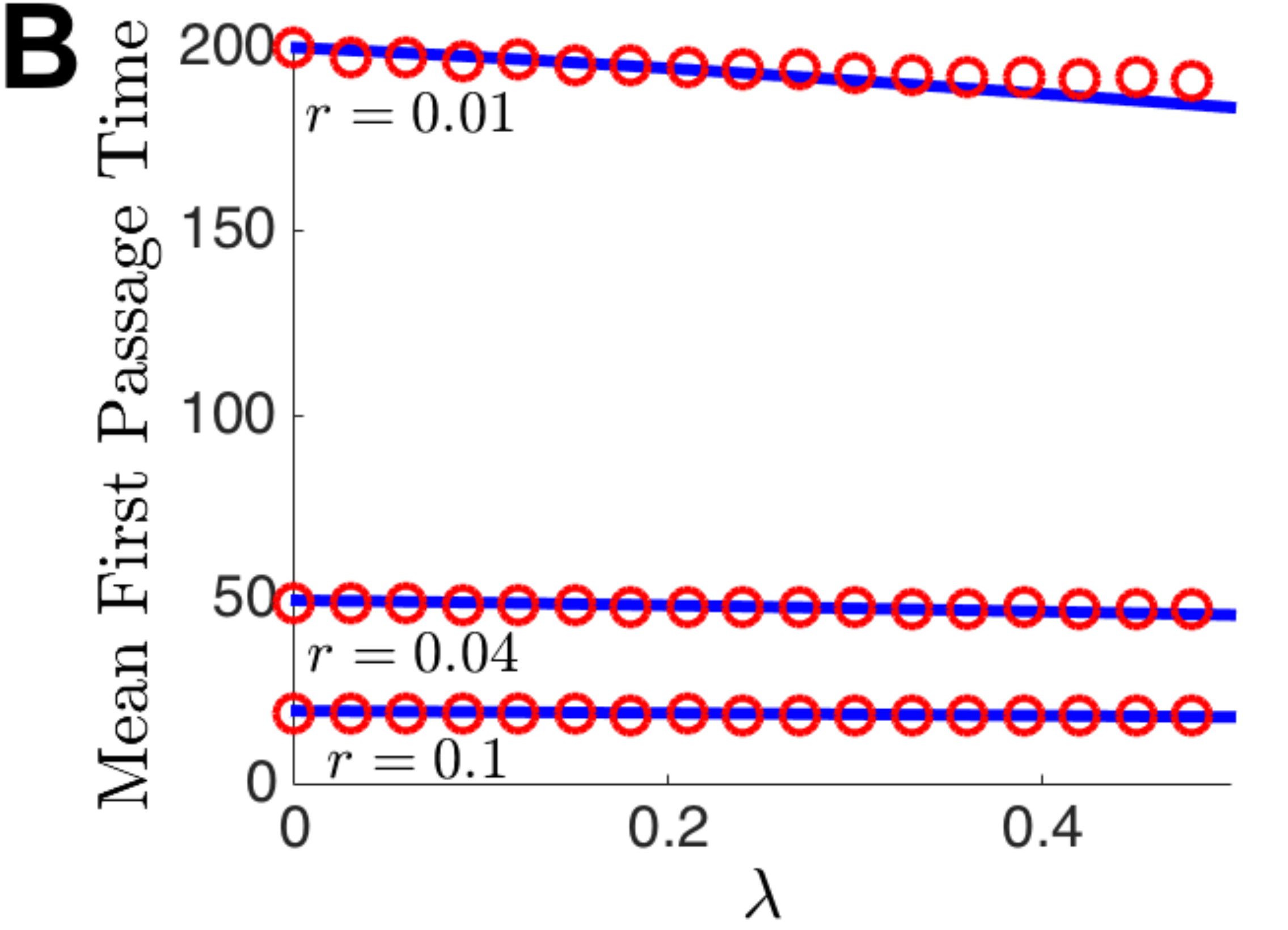} \\ \includegraphics[width=5.5cm]{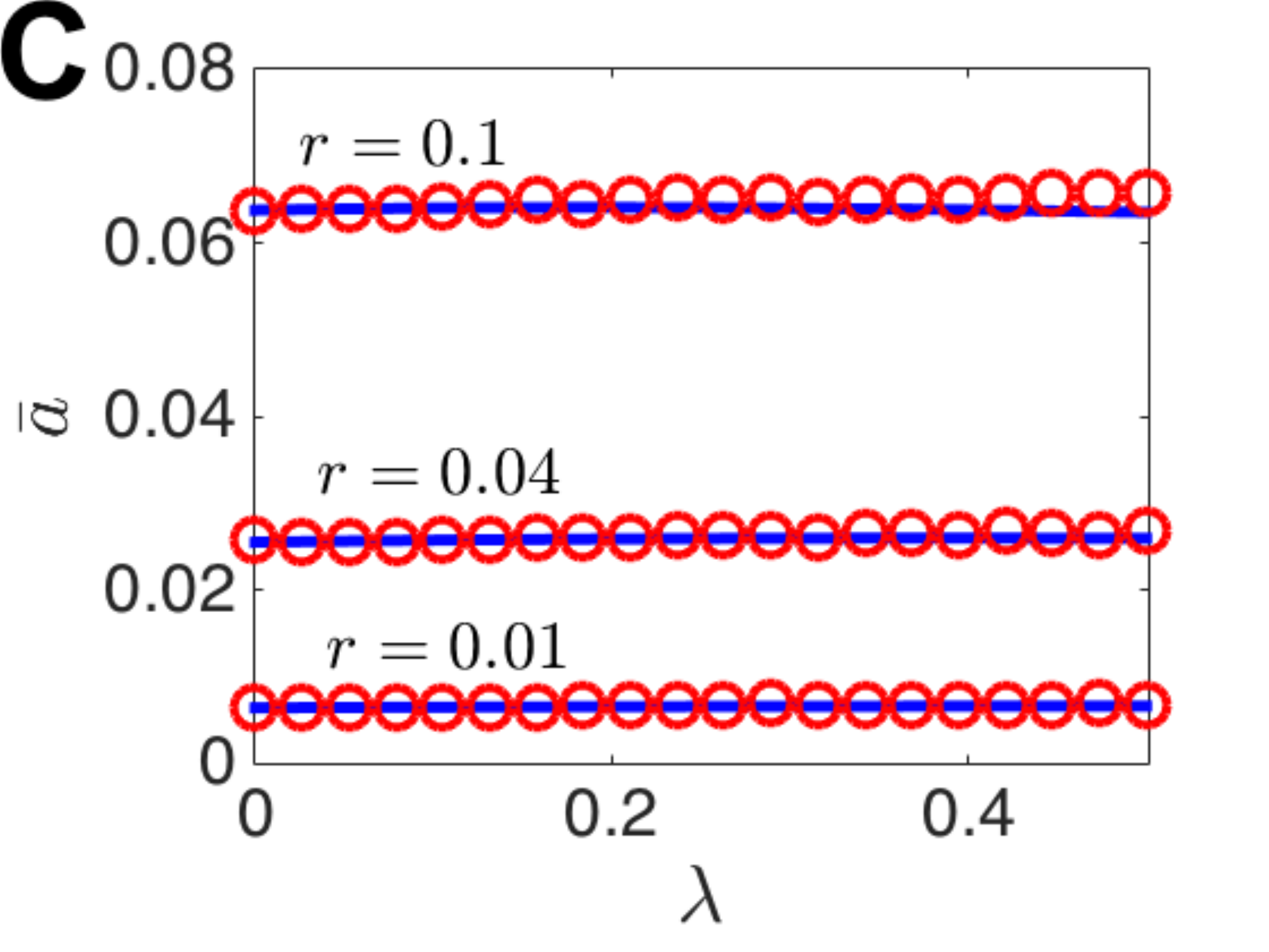} \includegraphics[width=5.5cm]{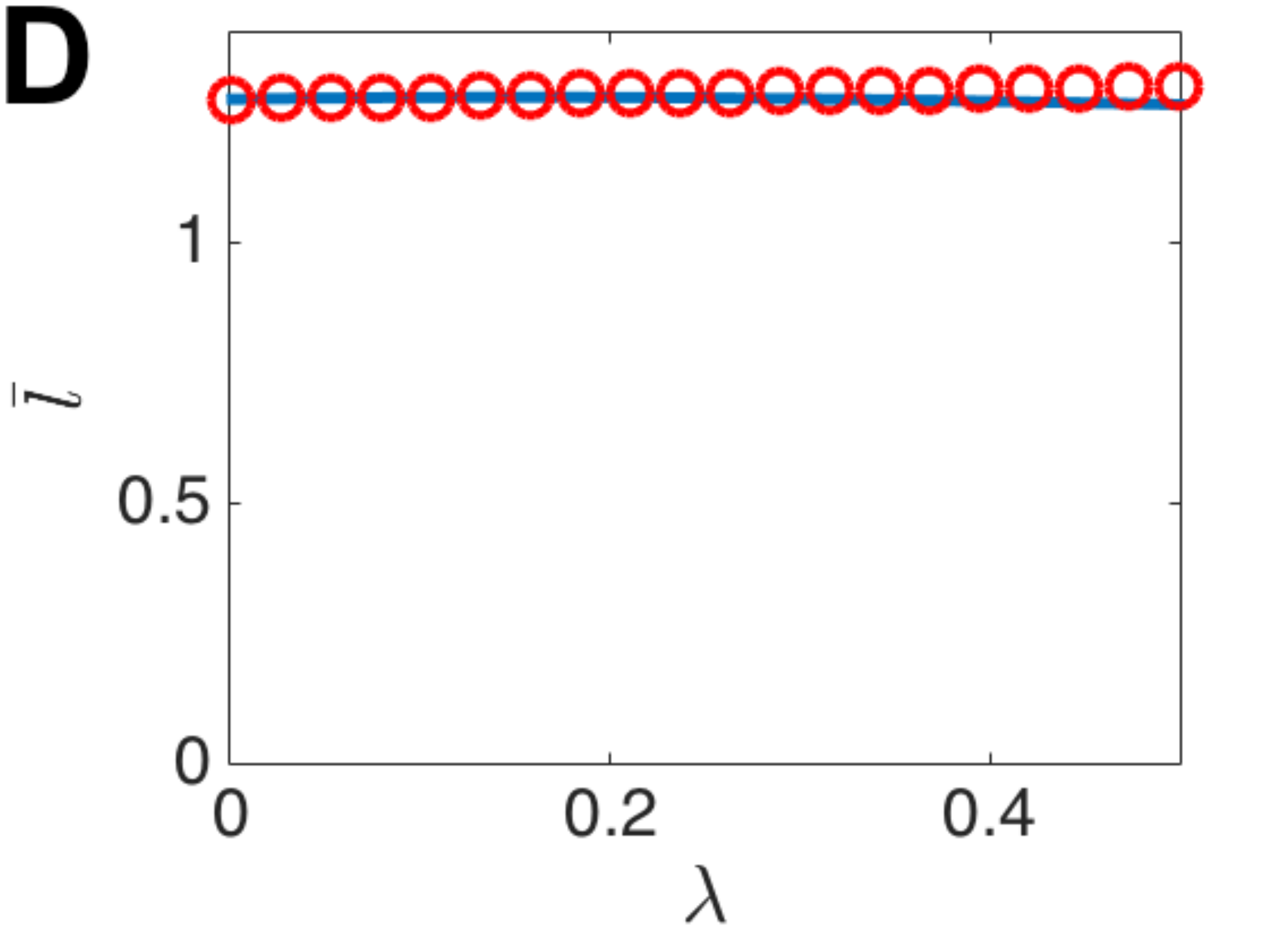} \end{center}
\caption{MFPT decreases when the searcher turns in the interior ($\lambda >0$). (A) To approximate the hitting probability $\bar{a} (\lambda)$ over a single path, we account for the new probability of hitting after turning $\alpha ( l, \theta )/ \pi $. (B) MFPT decreases as a function of $\lambda$, as demonstrated both by the theory (solid curve) in Eq.~(\ref{lambdahit}) and numerical simulations (circles). (C) Hitting probability for a single path between boundary points increases slightly with $\lambda$ according to theory Eq.~(\ref{lambdahit}) and numerical simulations. (D) Average path length also increases slightly with $\lambda$, as computed in Eq.~(\ref{lambdachord}). Note $f(z) = {\rm unif}[- \pi,\pi]$.}
\label{intturning}
\end{figure}

Once a single turn has been made, the probability of selecting a new direction which results in hitting the target is given by marginalizing across all possible locations of turns
\begin{align}
{\rm Pr}({\rm hit} | n = 1) \approx&  \frac{\lambda}{{\mc N} \pi } \int_{\alpha}^{\pi/2} \int_0^{2 \cos \theta} \e^{- \lambda x} \sin^{-1} \left[ \frac{r}{R \sqrt{1 + (x/R)^2 - 2 x \cos (\theta)/ R}} \right] d x d \theta \nonumber \\
& + \frac{\lambda}{{\mc N} \pi } \int_{0}^{\alpha} \int_0^{R-r} \e^{- \lambda x}\sin^{-1} \left[ \frac{r}{R \sqrt{1 + (x/R)^2 - 2 x \cos (\theta)/ R}} \right] d x d \theta \nonumber \\
= & {\mc H}(r,R),  \label{hitoneturn}
\end{align}
where $\alpha = \sin^{-1} (r/R)$ and the total normalization is given by integrating over the probability density $\lambda \e^{- \lambda x}$: ${\mc N} = \lambda \left( \int_{\alpha}^{\pi/2} \int_0^{2 \cos \theta}  \e^{- \lambda x} d x d \theta + \int_{0}^{\alpha} \int_0^{R-r} \e^{- \lambda x} d x d \theta \right)$.
Note that the angle of trajectories that will hit the target is larger for turns that occur  closer to the target (Fig. \ref{intturning}A). This may account for the slight increase in hitting probability due to turning (Fig. \ref{intturning}C).

Therefore, the total likelihood of hitting the target along a single path between boundary points can be linearly approximated by (a) subtracting the probability due to turning away from the target given by Eqs.~(\ref{noturn}) and (\ref{hitnoturn}) and (b) adding the probability due to turning towards the target as computed in Eqs.~(\ref{oneturn}) and (\ref{hitoneturn}):
\begin{align}
\bar{a} ( \lambda) = \e^{- \lambda (R-r)} \frac{2 \sin^{-1} (r/R)}{\pi} + \left( 1- \e^{- \lambda \overline{{\rm ch}}} \right)  \cdot {\mc H}(r,R).  \label{lambdahit}
\end{align}
This provides a new estimate for the probability of hitting in a single path, which we plot in Fig. \ref{intturning}C.

Furthermore, we can compute the average length of a single path between boundary points. This will no longer be given by the average chord length $\overline{{\rm ch}}$. Rather, our approximation will average in the paths consisting of two segments. Utilizing the probabilities of turning and not turning computed in Eq.~(\ref{noturn}) and (\ref{oneturn}), we can then appropriately weight the average lengths of one and two segment paths. First, note that paths with no turns will have an new length given by
\begin{align}
\bar{l}_0 = \frac{1}{{\mc N}_0} \int_0^{\pi/2} (2 \cos \theta) \e^{-2 \lambda \cos \theta} d \theta  \label{avnoturn}
\end{align}
with normalization constant ${\mc N}_0 = \int_0^{\pi/2} \e^{- 2 \lambda \cos \theta} d \theta$. Paths with a single turn will have length specified by their initial search angle $\theta$, first segment length $x$, and new angle $\phi$ following a turn. Given paths that start at $(x,y) = (R,0)$, the turning point will be $(x_0,y_0) = (R - x \cos \theta, x \sin \theta)$ and the new intersection point with the boundary will be
\begin{align*}
(x_c, y_c) = & \left( \sin \phi ( R \sin \phi  - x \sin(\theta + \phi) ) + \cos \phi \sqrt{R^2 - (x \sin ( \theta + \phi) - \sin \phi )^2}, \right. \\  & \left. \cos \phi ( x \sin(\theta + \phi) - R \sin \phi ) + \sin \phi \sqrt{R^2 - (x \sin ( \theta + \phi) - \sin \phi )^2} \right).
\end{align*}
For small targets ($r \ll R$), the effect of absorptions by the target will have a small effect on the average path length, so we marginalize over all three variables $\theta$, $x$, and $\phi$:
\begin{align}
\bar{l}_1 = \frac{\lambda}{{\mc N}_1} \int_0^{\pi/2} \int_0^{2 \cos \theta} \int_0^{2 \pi} \e^{- \lambda x} \left[ x + \sqrt{ (x_c - x_0)^2 + (y_c-y_0)^2 }\right] d \phi dx d \theta \label{avoneturn}
\end{align}
with normalization constant ${\mc N}_1 = 2 \pi \int_0^{\pi/2} 1-\e^{- 2 \lambda \cos \theta} d \theta$. Combining Eq.~(\ref{avoneturn}) with the average chord length, given no turns Eq.~(\ref{avnoturn}), we have the following estimate for the average path length
\begin{align}
\bar{l} ( \lambda)  = \e^{- \lambda \overline{{\rm ch}}} \bar{l}_0 + \left( 1- \e^{- \overline{{\rm ch}} \lambda} \right) \bar{l}_1, \label{lambdachord}
\end{align}
shown in Fig. \ref{intturning}D.

Incorporating Eqs.~(\ref{lambdahit}) and (\ref{lambdachord}) into Eq.~(\ref{binapp}), the formula for the MFPT, we can account for the effects of interior turning:
\begin{align}
{\mc T}_c(\lambda) = \bar{l} (\lambda) \left( \frac{1}{\bar{a}(\lambda)} - 1 \right) + R.
\end{align}
The main contribution to the reduction of the MFPT is due to a slight increase in the hitting probability $\bar{a}(\lambda)$ as shown in Fig. \ref{intturning}C. However, increasing turning $\lambda$ is does not significantly impact the time to find the target (Fig. \ref{intturning}B). Even for larger values of $\lambda$, the MFPT remains relatively unchanged as opposed to the case $\lambda = 0$.

\subsection{Spiral searches}
\label{spiralsect}

Both insects and mammals may utilize spiral patterned trajectories as search paths to locate a target~\cite{muller94,graziano03}. This can be more efficient and even optimal, since it can reduce the time spent in previously visited patches of the environment~\cite{alpern06}. However, spiral search may lead to unnecessarily long times needed to find the target if the spacing between rotations is too large or too small~\cite{reynolds07}.

\begin{figure}[t]
\begin{center} \includegraphics[width=7cm]{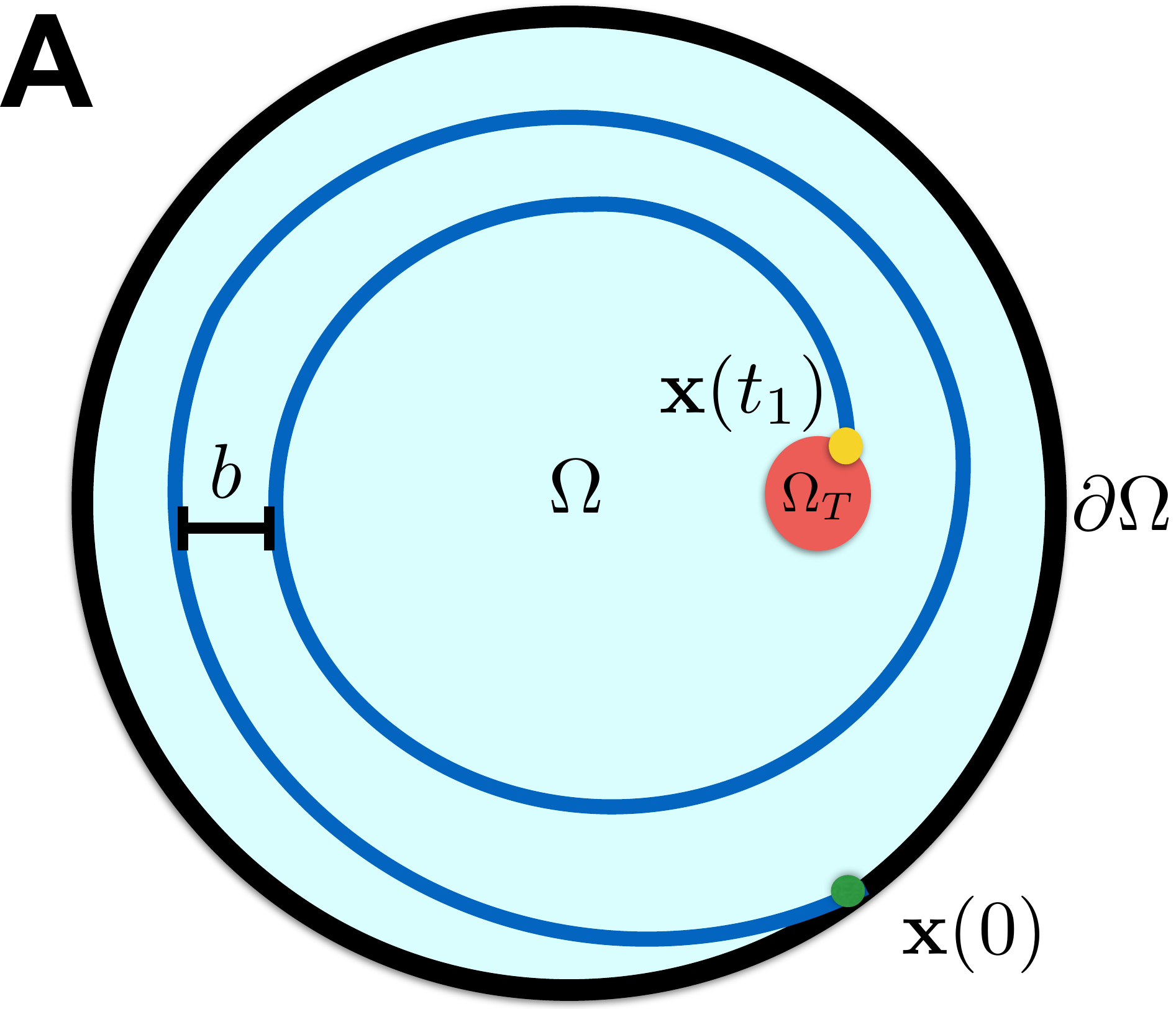}
\includegraphics[width=7cm]{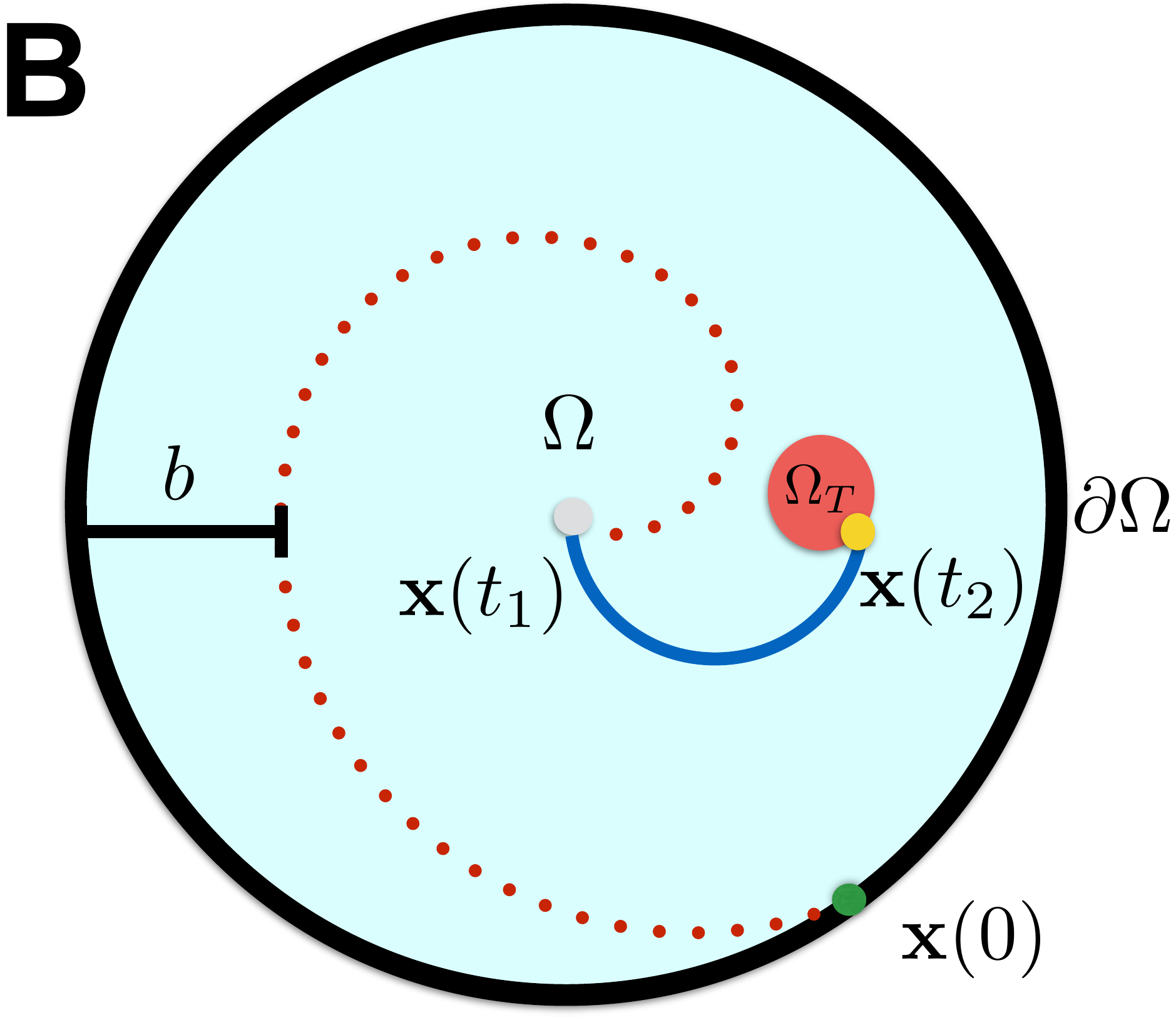} \\ 
\includegraphics[width=7cm]{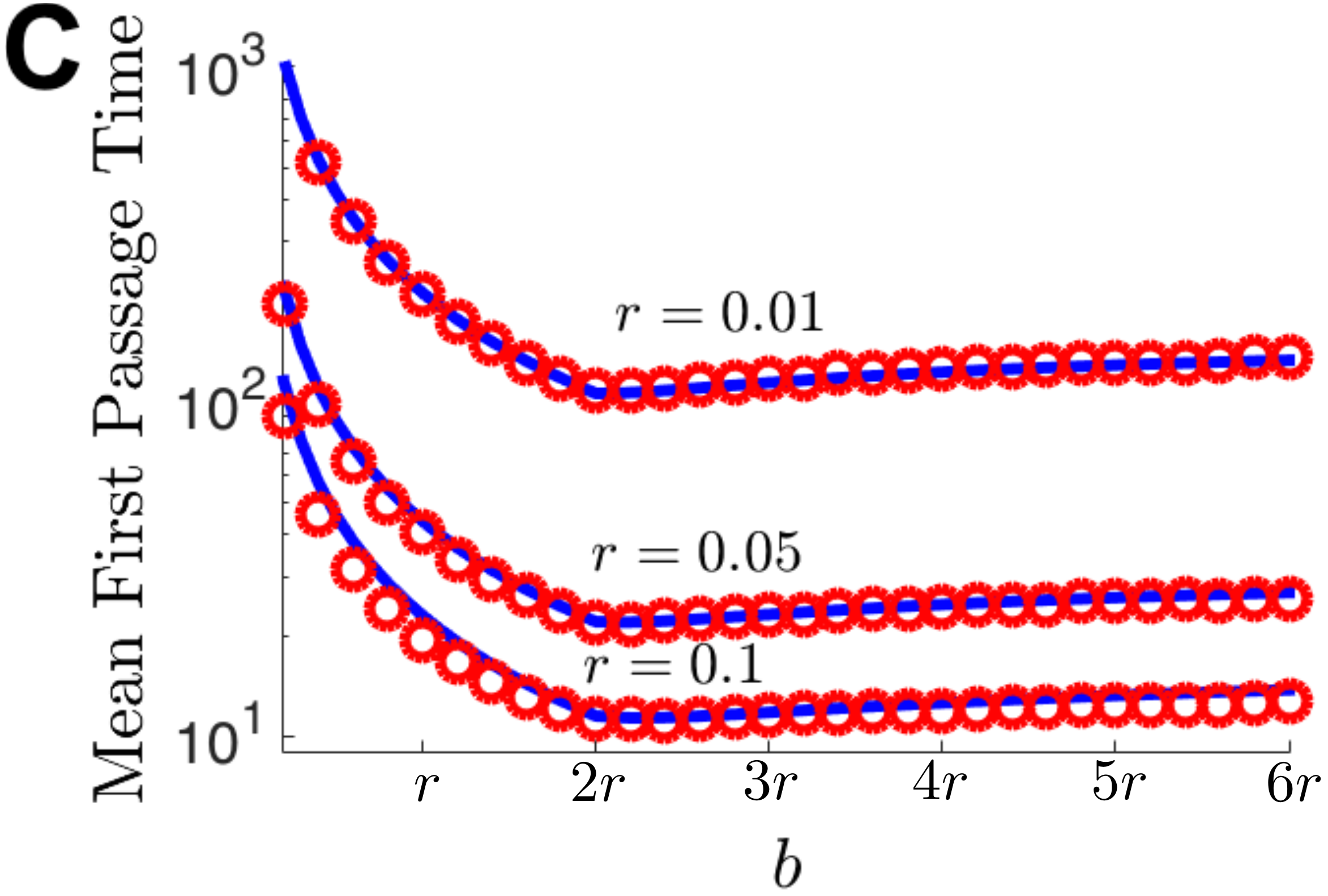}
\includegraphics[width=7cm]{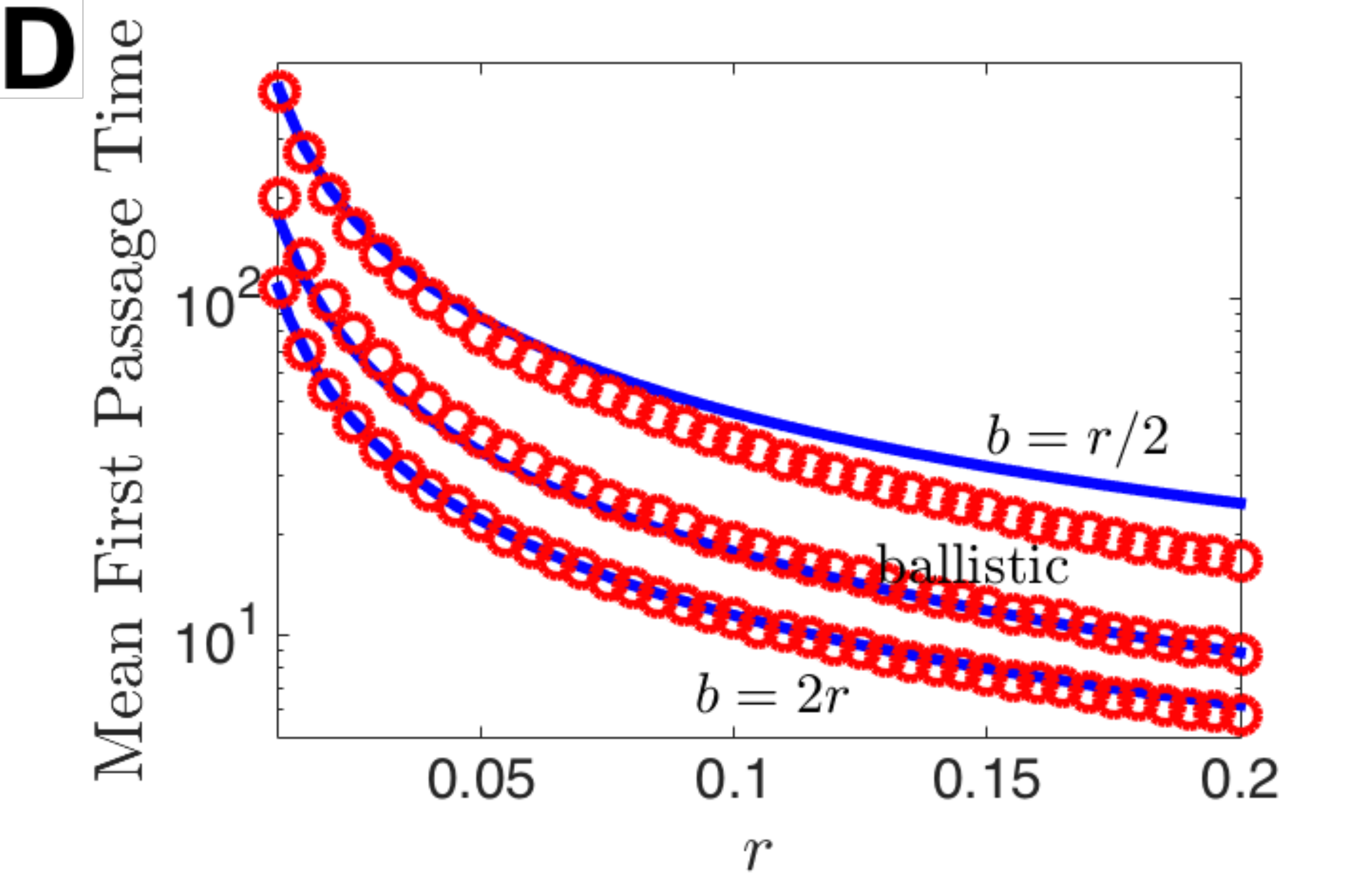} \end{center}
\caption{Spiral search strategies. (A) When the spiral width $b<2r$, the searcher finds the target in a single path. (B) When the spiral width $b>2r$, the searcher may miss the target, so a new spiral path begins at $(0,0)$. (C) MFPT to find the target is non-monotonic in the spiral width $b$, with a minimum at $b=2r$ the diameter of the target, as given by Eqs.~(\ref{spirepstime1}) and (\ref{spirepstime2}). (D) MFPT always decreases with target radius $r$. With an ideal spiral width $b=2r$, search takes less time than ballistic search, but search times can be longer than for ballistic search.
The probability of starting at angle $\theta$ on the boundary is uniform over $\theta \in [0,2\pi)$ as is the probability of drawing a search direction from the center. Domain has radius $R=1$.}
\label{spiralfig}
\end{figure}

We consider search trajectories described by an Archimedean spiral $\rho (\phi) = \frac{b}{2 \pi}\phi$ within the circular domain $\Omega$. Here $b$ is the closest distance between points along the same radius, effectively the width of the spiral. Were the radius of the target $r$ known to the searcher, the optimal coefficient could be chosen $b = 2 r$. This leads to no overlap in the environment searched while also ensuring that the target will be hit in a single search path. In cases where $b< 2 r$, the target is sure to be hit (Fig. \ref{spiralfig}A), but the searcher overlaps previously searched regions. If $b>2r$, the searcher may not hit the target during a single search path, since it will only hit targets with centers up to a distance $r$ away from its path (Fig. \ref{spiralfig}B). For uniformly randomly located targets, there will be a probability of approximately $\alpha = \frac{2r}{b}$ the target will be hit on a single uniformly randomly initiated spiral path. Unwrapping the spiral search path would reveal the searcher passes over an approximately rectangular strip of width $b$, but it can only spot targets whose centers are within a strip $2r$ about its path. This approximation worsens as the domain shrinks or the target size grows.

Any spiral search path that begins at angle $\phi_1$ on the interior and ends at $\phi_2$ on the periphery has length:
\begin{align}
L( \phi_1, \phi_2) &= \int_{\phi_1}^{\phi_2} \sqrt{ \rho(\phi)^2 + \rho'(\phi)^2 } d \phi = \frac{b}{2 \pi} \int_{\phi_1}^{\phi_2} \sqrt{\phi^2+1} d \phi = \frac{b}{2 \pi} \int_{\tan^{-1} \phi_1}^{\tan^{-1} \phi_2} \sec^3 u du \nonumber \\
&= \frac{b}{4 \pi} \left[ \phi_2 \sqrt{\phi_2^2 + 1} - \phi_1 \sqrt{\phi_1^2 + 1} + \ln \left| \frac{\sqrt{\phi_2^2 + 1} + \phi_2}{\sqrt{\phi_1^2+1} + \phi_1} \right| \right]. \label{spirlength}
\end{align}
For instance, in the case that $R \gg b$ and the search path has interior end point at $\phi_1 = 0$ and periphery end point $\phi_2 = \frac{2 \pi}{b} R$, we have
\begin{align*}
L( 0 , 2 \pi R/b) = \frac{R}{2} \sqrt{ \frac{4 \pi^2 R^2}{b^2} + 1} + \frac{b}{4 \pi}  \ln \left| \sqrt{ \frac{4 \pi^2 R^2}{b^2}  + 1} +  \frac{2 \pi R}{b} \right|  \approx \frac{\pi R^2}{b},
\end{align*}
so the circular domain can be approximately partitioned by the search path to form a rectangle of width $b$ and length $\pi R^2 / b$. 

We can apply the formula in Eq.~(\ref{spirlength}) to approximate the MFPT to find the target. Uniformly randomly choosing a location on the boundary $\theta \in [0, 2\pi)$, we consider a circular target of radius $r$ at a location $(\epsilon R,0)$ where $0 \leq \epsilon R \leq R-r$. For $b \leq 2r$, the time to find the target can be computed as ${\mc T}_s (\epsilon ) = L\left(  \frac{2\pi}{b}\epsilon R, \frac{2\pi}{b}R \right)$. Note, the MFPT will decrease as $b$ is increased to $2r$ as less time is wasted examining previously explored regions.  On the other hand, the searcher may choose a spiral width $b> 2r$, so $\alpha = \frac{2r}{b}$ approximates the probability of hitting the target in a single spiral. The MFPT to find the target is calculated by computing $h_j$ the likelihood of hitting the target after $j$ full spirals and $d_j$ the associated path length. The searcher starts at the periphery (at angle $\theta_2 =  \frac{2 \pi}{b} R$) and the target center is at angular location $\theta_1 = \frac{2 \pi}{b} \epsilon R$, so after $j$ full spiral paths the probability of locating the target is $h_j = \alpha (1-\alpha)^j$ with path-length:
\begin{align*}
d_j = j L(0,\theta_2) + \begin{cases} L(\theta_1,\theta_2) &j {\rm  even} \\ L(0,\theta_1) &j {\rm  odd} \end{cases}.
\end{align*}
Assembling these terms into an infinite sum, the average search path length is
\begin{align}
{\mc T}_s( \epsilon) = \sum_{j=0}^\infty h_j d_j = \frac{\displaystyle L \left( \frac{2 \pi \epsilon}{b}, \frac{2 \pi R}{b} \right) + (1-\alpha) L \left( 0,  \frac{2 \pi \epsilon}{b} \right)  }{2 - \alpha}  + \frac{\displaystyle (1 - \alpha) L \left( 0,  \frac{2 \pi R}{b} \right)}{\alpha},  \label{largebtime}
\end{align}
assuming the target is always fixed at location $(\epsilon R, 0)$.

So far, we have assumed the target location $(\epsilon R,0)$ to be fixed across realizations. However, targets may tend to appear randomly at any location in the domain $\Omega$. Using rotation symmetric, we project target locations to to $(\epsilon R,0)$ such that $\epsilon R \in [0, R-r]$.  Marginalizing in the case $b<2r$ we find ${\mc T}_s (\epsilon ) = L\left(  \frac{2\pi}{b}\epsilon R, \frac{2\pi}{b}R \right)$ and:
\begin{align}
\bar{\mc T}_s &= \frac{R}{R-r} \int_0^{1-r/R} L\left( \frac{2 \pi}{b} \epsilon R, \frac{2 \pi}{b}R \right) d \epsilon \nonumber \\
&= L \left(0, \frac{2 \pi}{b}R \right)  - \frac{b^2}{8 \pi^2 (R-r)} \int \limits_0^{2 \pi (R-r)/b} 
 x\sqrt{x^2 + 1}  + \ln|\sqrt{x^2+1} + x| dx \nonumber \\
&= L \left(0, \frac{2 \pi}{b}R \right) - F\left( \frac{2 \pi (R-r)}{b}\right) + F(0),
\label{spirepstime1}
\end{align}
where
\begin{align}
F(x) = \frac{b^2}{8 \pi^2 (R-r)} \bigg[ \frac{1}{3} (x^2 +1)^{3/2} + x \ln| \sqrt{x^2 + 1}+x| - \sqrt{x^2 +1} \bigg]. \label{Fspiral}
\end{align}
When $b>2r$, we average Eq.~(\ref{largebtime}) over the uniform density of $\epsilon R \in [0,R-r]$ so
\begin{align}
\bar{\mc T}_s =& \frac{R}{R-r} \int_0^{1-r/R} \frac{\displaystyle L \left( \frac{2 \pi \epsilon}{b}, \frac{2 \pi R}{b} \right) + (1-\alpha) L \left( 0,  \frac{2 \pi \epsilon}{b} \right)  }{2 - \alpha}  + \frac{\displaystyle (1- \alpha) L \left( 0,  \frac{2 \pi R}{b} \right)}{\alpha} d \epsilon \nonumber \\
= & \frac{\displaystyle  L \left(0, \frac{2 \pi}{b}R \right) - \alpha \left[ F\left( \frac{2 \pi (R-r)}{b}\right) - F(0) \right] }{2 - \alpha} + \frac{\displaystyle (1- \alpha) L \left( 0,  \frac{2 \pi R}{b} \right)}{\alpha}.  \label{spirepstime2}
\end{align}
We compare the theoretical approximations to the results from numerical simulations in Fig.\ref{spiralfig}C,D. As a function of $b$, the MFPT given by Eqs.~(\ref{spirepstime1}) and (\ref{spirepstime2}) are non-monotonic with a minimum at $b=2r$ (Fig. \ref{spiralfig}C).

\section{Ballistic search for multiple subdomains}
\label{multidomain}

\begin{figure}[t]
\begin{center} \includegraphics[width=10cm]{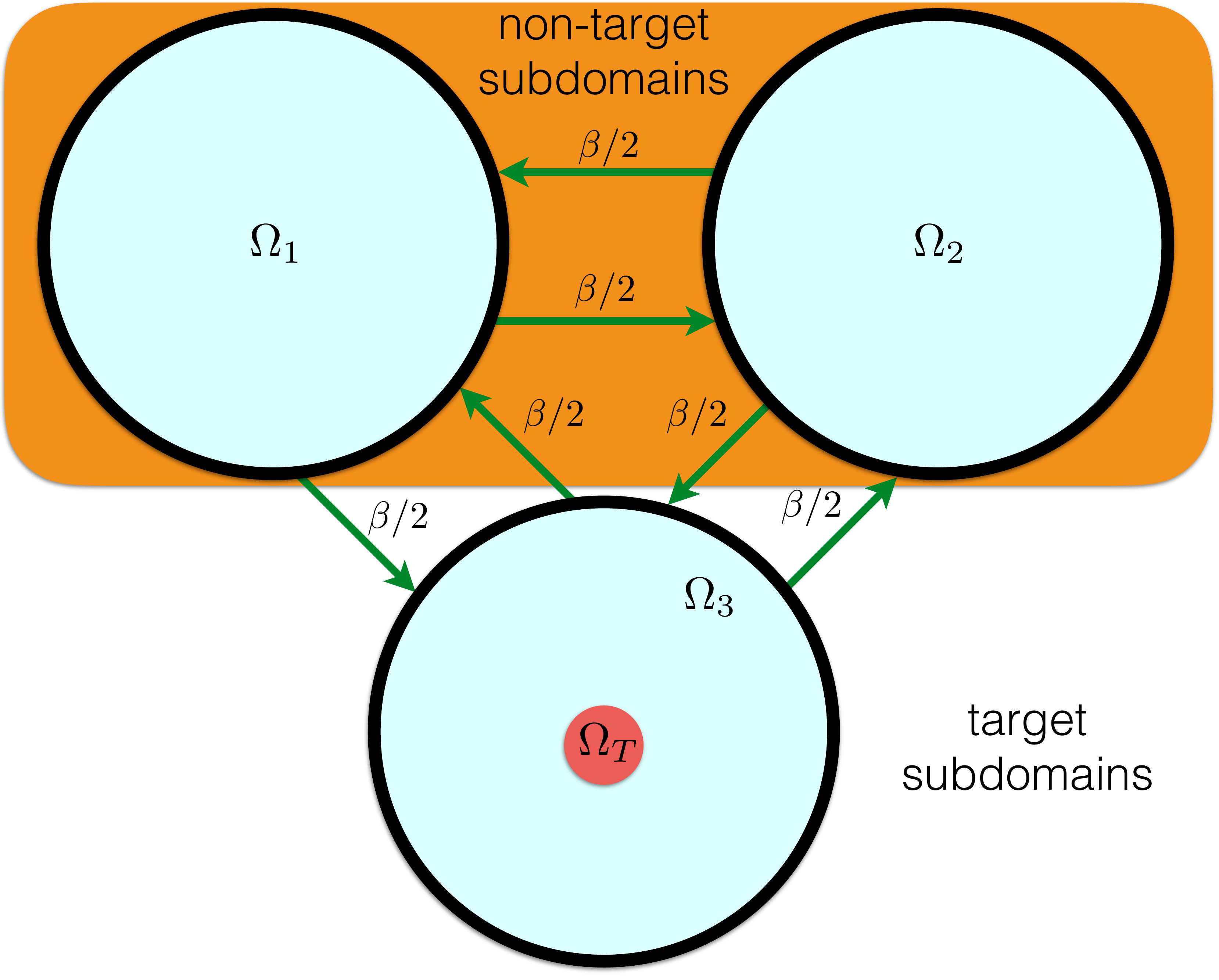} \end{center}
\caption{Domain with $N=3$ disconnected subdomains with a single target in subdomain $\Omega_3$. The total rate of transition out of each subdomain is $\beta$, and there is an equal likelihood of transitioning to one of the other two subdomains. The searcher can only locate the target when it is in subdomain $\Omega_3$. Transitions between subdomains take time $\langle T\rangle$ on average.}
\label{switchdiagram}
\end{figure}

Thus far, we have focused on search problems in a single bounded domain, but often organisms forage or search for shelter in patchy domains without searching between patches~\cite{krummel87,thompson01}. If patches are close together, an organism may frequently move between them to maximize search coverage of the environment. However, if patches are further apart, the organism may dwell in single patches for longer before moving on to search elsewhere. In light of this, we extend our analysis to the case of multiple disconnected subdomains $\Omega_j$ ($j=1,...,N$), where there is a mean travel time $\langle T \rangle$ between subdomains. There is only one target, located in the center of single subdomain $\Omega_k$ such that $\Omega_T \subset \Omega_k$, so searching in non-target subdomains ($\Omega_j$, $j \neq k$) will not yield a target hit. Thus, we introduce another free parameter into the strategy of our searcher $\beta$, the rate the searcher leaves its current subdomain $\Omega_j \mapsto \Omega_l$ ($l \neq j$). The searcher may only depart at subdomain boundaries $\partial \Omega_j$, and we assume it enters one of the other subdomains with equal probability $1/(N-1)$ (Fig. \ref{switchdiagram}). In our analysis, we aim to identify the optimal strategy for searching the disconnected domain for the single target, especially as it relates to the domain-switching rate $\beta$.

We now compute the MFPT to find the target of radius $r$ in a multiple subdomain environment $\Omega_1 \cup \Omega_2 \cup \cdots \cup \Omega_N$ where each subdomain is a circle of identical radius $R$. Assuming the searcher chooses a uniformly distributed random search angle, the likelihood of hitting the target along a single ballistic path, when inside the subdomain $\Omega_k$ is $\bar{a} = 2 \sin^{-1} (r/R)/\pi$. If the searcher is not in the subdomain with the target, the likelihood of hitting the target is zero. The probability of transitioning out of the target subdomain $\Omega_k$ before hitting the target $1 - \zeta$ ($\zeta$: probability of hitting target) is then
\begin{align}
1 - \zeta = \sum_{j=0}^{\infty} \beta (1-\bar{a})^{j+1} (1- \beta)^j = \frac{\beta (1- \bar{a})}{\bar{a} + \beta (1-\bar{a})}  \implies \zeta = \frac{\bar{a}}{\bar{a} + \beta(1- \bar{a})}.  \label{zetamult}
\end{align}
Once the searcher leaves the target subdomain, it may transition between several non-target subdomains before returning. Note the probability of transitioning from a non-target subdomain to another non-target subdomain is $\xi = (N-2)/(N-1)$. Thus, the average number of non-target subdomains visited before returning to $\Omega_k$ is
\begin{align}
\sum_{j=0}^{\infty} (j+1) \xi^j (1- \xi) = N-1.  \label{numbnontar}
\end{align}
To determine the average amount of time spent searching, we need to compute the average time spent searching per visit to the target subdomain when finding the target
\begin{align*}
{\mc T}_{tf} = \frac{\overline{{\rm ch}}}{\zeta} \sum_{j=0}^{\infty} j \bar{a} (1- \bar{a})^j (1 - \beta)^j + R-r = \frac{\overline{{\rm ch}} (1- \bar{a}) (1- \beta)}{\bar{a} + \beta (1 - \bar{a})} + R-r,
\end{align*}
and not finding the target
\begin{align*}
{\mc T}_{tl} =  \frac{\overline{{\rm ch}}}{1 - \zeta} \sum_{j=0}^{\infty} (j+1) \beta (1- \bar{a})^{j+1} (1 - \beta)^j = \frac{\overline{{\rm ch}}}{\bar{a} + \beta (1- \bar{a})}.
\end{align*}
Furthermore, the average amount of time spent in a single non-target subdomain before departing is
\begin{align*}
{\mc T}_{n} = \overline{{\rm ch}} \sum_{j=0}^{\infty} (j+1) \beta (1- \beta)^j = \frac{ \overline{{\rm ch}}}{\beta},
\end{align*}
so along with Eq.~(\ref{numbnontar}), the average time between trips to the target subdomain is
\begin{align*}
{\mc T}_{na} =  \frac{(N-1) \overline{{\rm ch}}}{\beta}.
\end{align*}
We pair these times along with the probability of hitting the target on the $j$th visit to the target domain. For the time being, we focus on the case where $\langle T \rangle = 0$, so in the case where the searcher begins in the target subdomain, the time to find the target is
\begin{align*}
{\mc T}_t &= \sum_{j=0}^{\infty} \zeta (1 - \zeta)^j \left[ j {\mc T}_{na} + j {\mc T}_{tl} + {\mc T}_{tf} \right] = \frac{1 - \zeta}{\zeta} \left[ {\mc T}_{na} + {\mc T}_{tl} \right] + {\mc T}_{tf} \\
&= \frac{\beta(1- \bar{a})}{\bar{a}} \left[  \frac{(N-1) \overline{{\rm ch}}}{\beta} + \frac{\overline{{\rm ch}}}{\bar{a} + \beta (1- \bar{a})} \right] + \frac{\overline{{\rm ch}} (1- \bar{a}) (1- \beta)}{\bar{a} + \beta (1 - \bar{a})} + R-r \\
&= N \overline{{\rm ch}} \left( \frac{1}{\bar{a}} -1 \right)  + R-r.
\end{align*}
When the searcher begins in a non-target subdomain, the time to find the target is
\begin{align*}
{\mc T}_o &= \sum_{j=0}^{\infty} \zeta (1 - \zeta)^j \left[ (j+1) {\mc T}_{na} + j {\mc T}_{tl} + {\mc T}_{tf} \right] = \frac{1 - \zeta}{\zeta} \left[ {\mc T}_{na} + {\mc T}_{tl} \right] +  {\mc T}_{na} + {\mc T}_{tf} \\
&= \frac{\beta(1- \bar{a})}{\bar{a}} \left[  \frac{(N-1) \overline{{\rm ch}}}{\beta} + \frac{\overline{{\rm ch}}}{\bar{a} + \beta (1- \bar{a})} \right] +  \frac{(N-1) \overline{{\rm ch}}}{\beta} +  \frac{\overline{{\rm ch}} (1- \bar{a}) (1- \beta)}{\bar{a} + \beta (1 - \bar{a})} + R-r   \\
&= N \overline{{\rm ch}} \left( \frac{1}{\bar{a}} -1 \right) +  \frac{(N-1) \overline{{\rm ch}}}{\beta}   + R-r.
\end{align*}
If initial conditions are uniformly distributed across subdomains, the probability of starting in the target subdomain is $1/N$, so the generalized MFPT is
\begin{align}
{\mc T}_c = \frac{1}{N} {\mc T}_t + \frac{N-1}{N} {\mc T}_o = N \overline{{\rm ch}} \left[ \left( \frac{1}{\bar{a}} -1 \right) +  \frac{(N-1)^2}{N^2 \beta } \right] + R-r.  \label{mfptnotransit}
\end{align}
In the case of instantaneous transits between subdomains $T = 0$, the best strategy is to transition at every boundary encounter, $\beta = 1$, allowing rapid coverage of the domain. This is due to the fact that the searcher has no knowledge of the location of target until it has been located, and transitions proceed randomly between subdomains.

\begin{figure}[t]
\begin{center} \includegraphics[width=6.5cm]{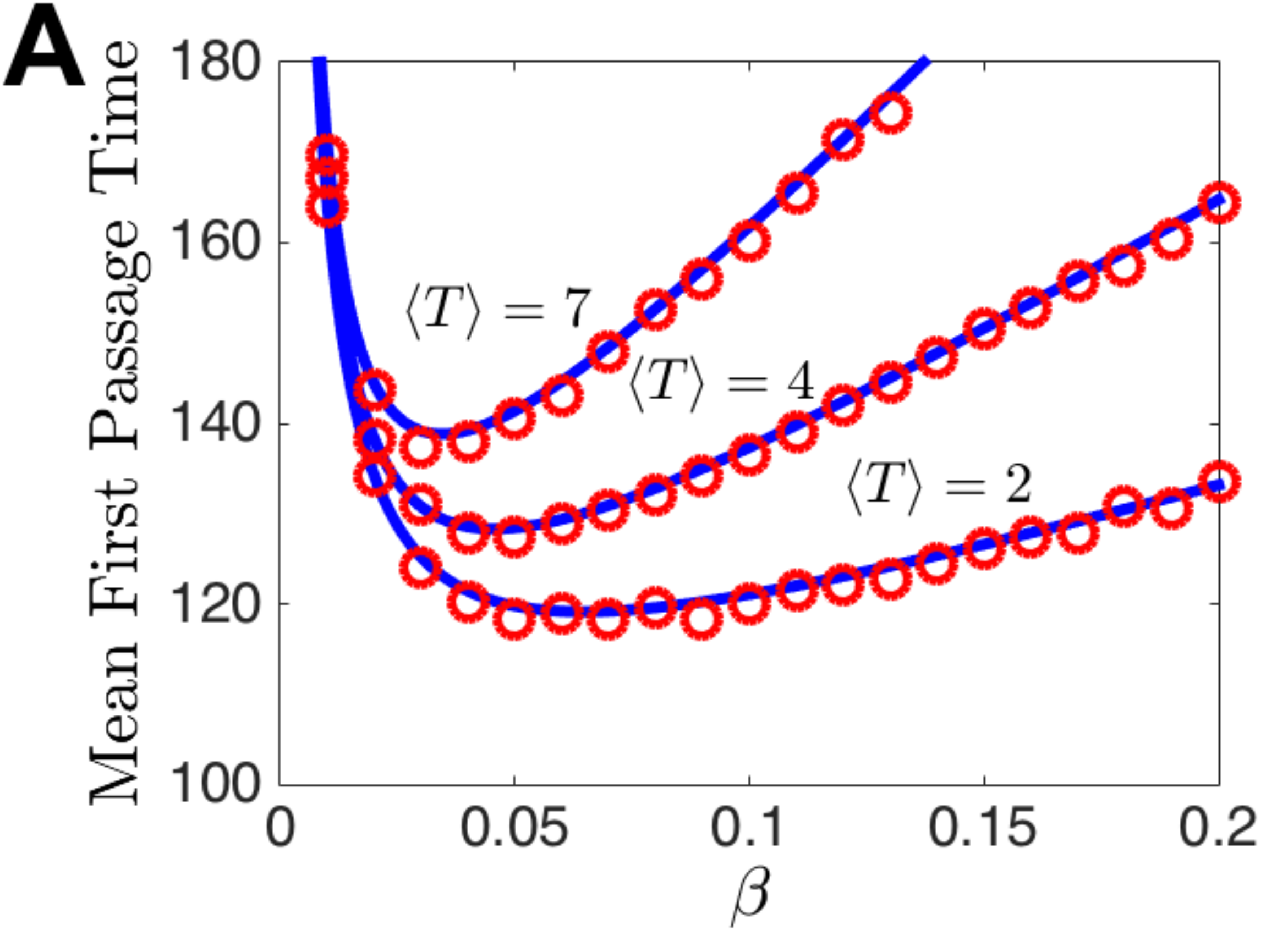} \includegraphics[width=6.5cm]{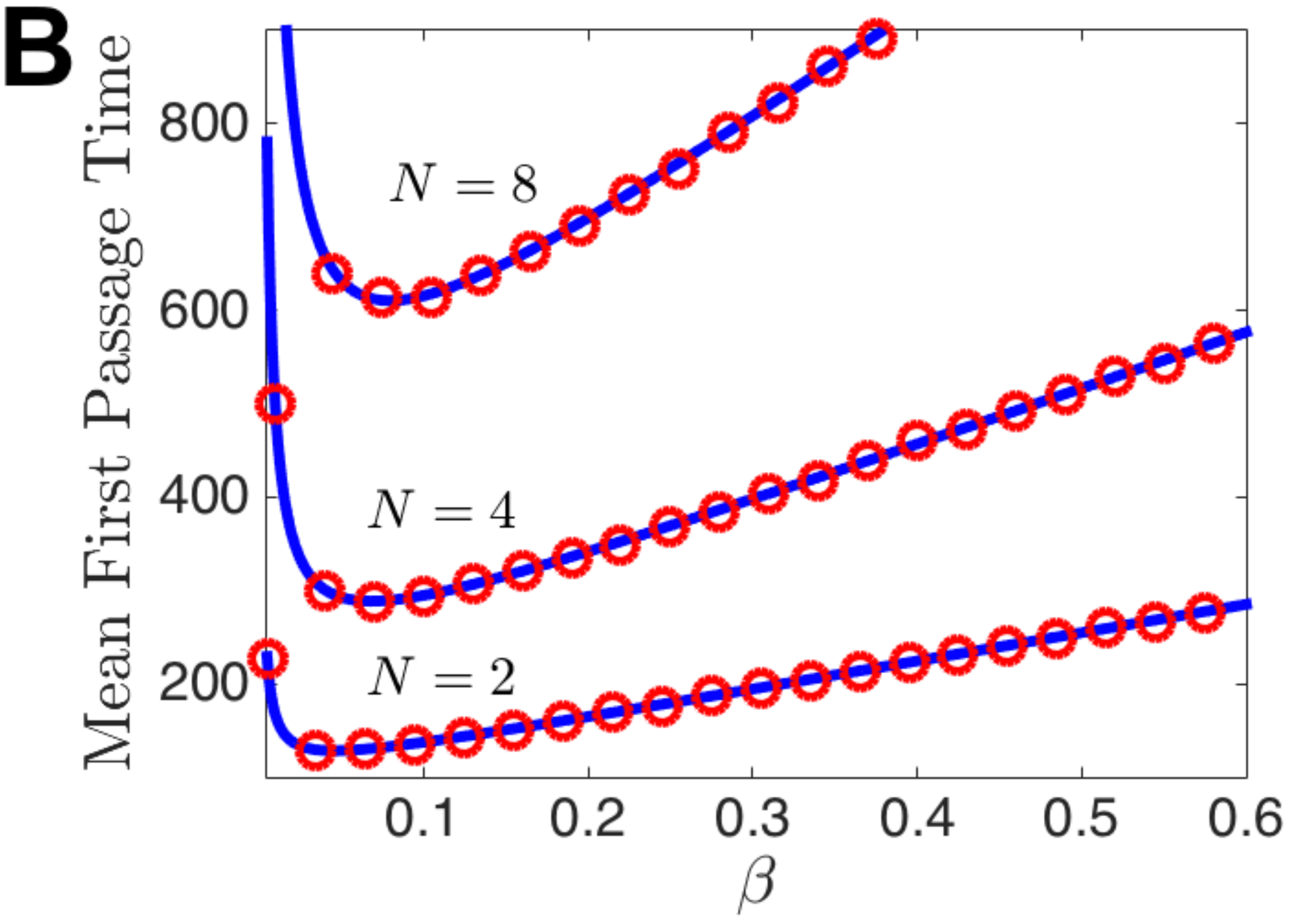}  \end{center}
\caption{Minimizing the MFPT by varying the transition rate $\beta$. (A) In a dual subdomain ($N=2$) environment, the MFPT is non-monotonic in $\beta$, obtaining a minimum at the value given by Eq.~(\ref{betmin}). Numerical simulations (circles) match well with the theory (solid line) given by Eq.~(\ref{fullmfptmulti}). As the average transit time $\langle T \rangle$ is increased, the optimal transition rate $\beta_{{\rm min}}$ decreases. $\langle T \rangle$ was distributed according to a uniform distribution of length 4, a standard normal distribution centered at 4, and an exponential distribution translated 4 time units with scaling parameter 3, giving mean transit times $\langle T \rangle = 2,4,7$ respectively. (B) As the number of domains $N$ is increased, so does the MFPT. However, the optimal transition rate $\beta_{{\rm min}}$ remains roughly the same. We chose target radius is $r=0.04$ and the search direction at the boundary is $f(z) = {\rm unif}[- \pi, \pi]$. The mean transit time $\langle T \rangle$ was distributed normally with mean 4 and variance 1 for all domain curves.}
\label{multimfpt}
\end{figure}

Now, we introduce random transit times $T$ given by a rectified Gaussian distribution $T \sim {\mc N}^R( \mu, \sigma^2)$, capturing the variability possible in organisms time to move from patch to patch. Considering nonzero transit times $\langle T \rangle >0$ and how this augments the MFPT Eq.~(\ref{mfptnotransit}), we note that when the searcher begins in the target subdomain the additional time due to transit will be
\begin{align*}
{\mc P}_t &= \sum_{j=0}^{\infty} \zeta (1- \zeta)^j \left[ j N \langle T \rangle \right] = \beta N \langle T \rangle \left( \frac{1}{\bar{a}} - 1 \right),
\end{align*}
whereas when it begins in a non-target subdomain, the additional transit time is
\begin{align*}
{\mc P}_o &=  \sum_{j=0}^{\infty} \zeta (1- \zeta)^j \left[ j N \langle T \rangle  + (N-1) \langle T \rangle \right] =  \beta N \langle T \rangle \left( \frac{1}{\bar{a}} - 1 \right) + (N-1) \langle T \rangle,
\end{align*}
so the updated MFPT is
\begin{align}
{\mc T}_c (\beta) = N ( \overline{{\rm ch}} + \beta \langle T \rangle ) \left[ \left( \frac{1}{\bar{a}} -1 \right) +  \frac{(N-1)^2}{N^2 \beta } \right] + R-r.  \label{fullmfptmulti}
\end{align}
To identify the optimal switching rate $\beta = \beta_{{\rm min}}$ that minimizes the MFPT ${\mc T}_c(\beta)$, we differentiate
\begin{align*}
\frac{d {\mc T}_c(\beta)}{d \beta} = N \langle T \rangle \left( \frac{1}{\bar{a}} -1 \right) - \frac{\overline{{\rm ch}}(N-1)^2}{N \beta^2},
\end{align*}
and note ${\mc T}_c''(\beta) = 2 \overline{{\rm ch}}(N-1)^2 /(N\beta^3)>0$ for $\beta>0$. Thus, any critical points occurring when $\beta>0$ are minima. To identify the minimum, we set ${\mc T}_c'(\beta_{{\rm min}}) = 0$ and solve for
\begin{align}
\beta_{{\rm min}} = \frac{N-1}{N} \sqrt{ \frac{\overline{{\rm ch}} \bar{a}}{\langle T \rangle (1 - \bar{a})} }. \label{betmin}
\end{align}
Thus, the optimal switching rate $\beta_{{\rm min}}$ is inversely proportional to $\sqrt{\langle T \rangle}$, so the switching rate should decrease as the transit time increases (Fig. \ref{multimfpt}A). This allows for a more thorough search of a single subdomain before transitioning. Interestingly, Eq.~(\ref{betmin}) is roughly constant in the variable $N$ as it is increased (Fig. \ref{multimfpt}B). Thus, even for a very large number of domains $N \gg 1$, the parameters that determine the best switching rate are the chord length $\overline{{\rm ch}}$, probability of hitting the target $\bar{a}$, and the transition time $\langle T \rangle$.

\section{Discussion}

We have studied a velocity-jump process model of persistent search in bounded domains. Initially, we considered a searcher that only turned on the domain boundary. Paths of the searcher are partitioned into segments that link points on the boundary. To derive the average MFPT to find the target, we approximated the average probability of hitting the target in a single segment $\bar{a}$. Pairing this with our approximation of an average segment-length $\overline{{\rm ch}}$, we then marginalized over all possible search path lengths. Importantly, we modeled the search process as memoryless, so each segment was assumed to have been drawn from the same distribution. Applying this to single domains, we found the time to find the target decreases for targets closer to the boundary. When searchers had a small probability of turning on the interior of the domain, the time to find the target decreases slightly, due to an increase in the hitting probability of a single segment. Lastly, in domains comprised of multiple disconnected subdomains, a key parameter in determining the optimal search strategy is the time it takes the searcher to move between subdomains. Ultimately, we found the searcher should move between domains less often when subdomain transitions take longer.

Our study provides an idealized model of an organism's search for a target in a confined domain. This model could apply to animals foraging in a patchy environment~\cite{thompson01} or looking for shelter in controlled experiments~\cite{jeanson03} as well as their natural habitat~\cite{reynolds07}. Furthermore, the velocity-jump process can produce long spatial correlations~\cite{othmer88,codling08}, similar to those often observed in statistical analyses of organismal motion~\cite{bergman00,bartumeus05}. Our analysis revealed that low-probability ($\lambda \ll 1$) turning on the interior of a bounded domain can lead to a decrease in the time to find the target. It would be interesting to develop theory for analyzing the case in which there is a high probability of turning ($\lambda \gg 1$). In this limit, the velocity-jump process can be approximated by a diffusion process~\cite{othmer00}, so it may be possible to perturb around this limit to approximate the effects of lengthening the spatial correlations in random walk segments. Theory for diffusion to a small target in planar domains is well understood, so we could leverage some of these previous results~\cite{holcman04,condamin07,pillay10}.

Also, the theory we developed for the case of multiple disconnected subdomains could be extended to other search processes. For instance, intermittent search processes with non-reactive and reactive phases are a better model of foraging processes in some situations~\cite{benichou11}. In this case, we could still separate the search process into time spent in non-target and target domains. The main difference would be that the likelihood of hitting the target within the target domain be computed in the case where the searcher is intermittently reactive. Similar extensions could be applied to diffusive search. Time spent in each subdomain would then be characterized by finding the mean time for the searcher to complete a random walk that starts and ends at the boundary.

\section*{Acknowledgements}

This work was supported by NSF grants (DMS-1517629 and DMS-1311755).

\section*{References}
\bibliographystyle{iopart-num}
\providecommand{\newblock}{}

\end{document}